\documentclass[letterpaper,twocolumn,prb,aps,showpacs,superscriptaddress,floatfix]{revtex4-1}
\usepackage[latin1]{inputenc}
\usepackage{bm}
\usepackage[usenames]{color}
\usepackage{multirow}
\usepackage{amssymb}
\usepackage{amsbsy}
\usepackage{amsmath}
\usepackage{stmaryrd}
\usepackage{graphicx}
\usepackage{epsfig}
\usepackage{placeins}
\usepackage{ulem}
\usepackage{soul}
\makeatletter

\begin{document}

\title{Spin and energy currents in integrable and nonintegrable spin-$1/2$ chains:\\
A typicality approach to real-time autocorrelations}

\author{Robin Steinigeweg}
\email{r.steinigeweg@tu-bs.de} \affiliation{Institute for Theoretical
Physics, Technical University Braunschweig, D-38106 Braunschweig,
Germany}

\author{Jochen Gemmer}
\email{jgemmer@uos.de} \affiliation{Department of Physics,
University of Osnabr\"uck, D-49069 Osnabr\"uck, Germany}

\author{Wolfram Brenig}
\email{w.brenig@tu-bs.de} \affiliation{Institute for Theoretical
Physics, Technical University Braunschweig, D-38106 Braunschweig,
Germany}

\date{\today}

\begin{abstract}
We use the concept of typicality to study the real-time dynamics of
spin and energy currents in spin-$1/2$ models in one dimension and
at nonzero temperatures. These chains are the integrable XXZ chain
and a nonintegrable modification due to the presence of a staggered
magnetic field oriented in $z$ direction. In the framework of linear
response theory, we numerically calculate autocorrelation functions
by propagating a single pure state, drawn at random as a typical
representative of the full statistical ensemble. By comparing to
small-system data from exact diagonalization (ED) and existing
short-time data from time-dependent density matrix renormalization
group (tDMRG), we show that typicality is satisfied in finite
systems over a wide range of temperature and is fulfilled in both,
integrable and nonintegrable systems. For the integrable case, we
calculate the long-time dynamics of the spin current and extract the
spin Drude weight for large systems outside the range of ED. We
particularly provide strong evidence that the high-temperature Drude
weight vanishes at the isotropic point. For the nonintegrable case,
we obtain the full relaxation curve of the energy current and
determine the heat conductivity as a function of magnetic field,
exchange anisotropy, and temperature.
\end{abstract}

\pacs{05.60.Gg, 71.27.+a, 75.10.Jm}

\maketitle

\section{Introduction}

The concept of typicality \cite{gemmer2003, goldstein2006,
reimann2007, popescu2006, white2009, hams2000, bartsch2009,
sugiura2012, elsayed2013, steinigeweg2014-1, steinigeweg2014-2}
states that a single pure state can have the same properties as the
full statistical ensemble. This concept is not restricted to
specific states and applies to the overwhelming majority of all
possible states, drawn at random from a high-dimensional Hilbert
space. In the cleanest realization, even a single eigenstate of the
Hamiltonian may feature the properties of the full equilibrium
density matrix, assumed in the well-known eigenstate thermalization
hypothesis \cite{deutsch1991, srednicki1994, rigol2008}. The notion
of property is manifold in this context and also refers to the
expectation values of observables. Remarkably, typicality is not
only a static concept and includes the dynamics of expectation
values \cite{bartsch2009}. Recently, it has become clear that
typicality even provides the basis for powerful numerical approaches
to the dynamics of quantum many-particle systems at nonzero
temperatures \cite{hams2000, elsayed2013, steinigeweg2014-1,
steinigeweg2014-2}. These approaches are in the center of this
paper.

Understanding relaxation and transport dynamics in quantum many-body
systems is certainly one of the most desired and ambitious aims of
condensed-matter physics and experiencing an upsurge of interest in
recent years, both experimentally and theoretically. On the one
hand, the advent of ultracold atomic gases raises challenging
questions about the equilibration and thermalization in isolated
many-particle systems \cite{cazalilla2010}, including the existence,
origin, or speed of relaxation processes in the absence of any
external bath. On the other hand, future information technologies
such as spintronics call for a deeper insight into transport
dynamics of quantum degrees of freedom such as spin excitations.
Spin transport in conventional nano-systems \cite{appelbaum2007,
tombros2007, stern2008, kuemmeth2008} is inevitably linked to the
dynamics of itinerant charge carriers. In contrast, Mott-insulating
quantum magnets allow for pure spin currents and thus open new
perspectives in quantum transport. In the past decade, magnetic
transport in one-dimensional quantum magnets has attracted
considerable attention because of the discovery of very large
magnetic heat-conduction \cite{sologubenko2000, hess2001,
hlubek2010} and long nuclear magnetic relaxation times
\cite{thurber2001, kuehne2010}. Genuine spin transport, however,
still remains to be observed in experiments and particularly its
classification in terms of ballistic or diffusive propagation is an
issue of ongoing experimental research \cite{maeter2012, xiao2014}.

The theoretical study of transport in low-dimensional quantum
magnets has a long and fertile history. Amongst all questions, the
dissipation of currents is a key issue and has been investigated
extensively at zero momentum and frequency in connection with the
linear-response Drude weight \cite{zotos1997, kluemper2002,
shastry1990, zotos1999, benz2005, prosen2011, prosen2013,
narozhny1998, heidrichmeisner2003, heidrichmeisner2007,
herbrych2011, steinigeweg2013, fujimoto2003}. The Drude weight is
the nondissipating part of the current autocorrelation function and,
if existent, indicates a ballistic channel close to equilibrium.
While in generic nonintegrable systems it is commonly expected that
Drude weights do not exist in the thermodynamic limit
\cite{heidrichmeisner2003}, the picture is different and more
complicated in integrable systems: Since an overlap of currents with
the macroscopic number of conserved quantities is probable, Drude
weights are expected to exist. But this overlap is not necessarily
finite for all model parameters and Drude weights can vanish in the
thermodynamic limit. In this context, an important example is the
antiferromagnetic and anisotropic Heisenberg (XXZ) spin-$1/2$ chain.
This chain is studied in the present paper. It is a fundamental
model for the magnetic properties of interacting electrons in low
dimensions. It is not only relevant to the physics of
one-dimensional quantum magnets \cite{johnston2000} but also to
physical questions in a much broader context \cite{trotzky2007,
gambardella2006, kruczenski2004, kim1996}.

In the XXZ spin-$1/2$ chain, the heat current is strictly conserved
for all values of the exchange anisotropy $\Delta$ \cite{zotos1997,
kluemper2002} and energy flows through a ballistic channel only.
This type of flow is at the heart of the colossal heat conduction
observed experimentally in almost ideal material realizations of the
model \cite{sologubenko2000, hess2001, hlubek2010}. In contrast, the
spin current is not strictly conserved and the existence of a spin
Drude weight is a demanding problem, resolved only partially despite
much effort. At zero temperature, $T = 0$, early work
\cite{shastry1990} showed that the spin Drude weight is nonzero in
the gapless regime $\Delta \leq 1$ (metal) but zero in the gapped
regime $\Delta > 1$ (insulator). Bethe-Ansatz solutions
\cite{zotos1999, benz2005} support a qualitatively similar picture
at nonzero temperatures, $T > 0$, but with a disagreement at the
isotropic point $\Delta = 1$. Recent progress in combining
quasi-local conservation laws and Mazur's inequality has lead to a
rigorous lower bound to the spin Drude weight in the limit of high
temperatures \cite{prosen2011, prosen2013}. This bound is very close
to the Bethe-Ansatz solution but still allows for a vanishing Drude
weight at $\Delta = 1$.

Numerically, a large variety of sophisticated methods has been
applied to transport and relaxation dynamics in the anisotropic
Heisenberg spin-$1/2$ chain, including full exact diagonalization
(ED) \cite{narozhny1998, heidrichmeisner2003, heidrichmeisner2007,
herbrych2011, steinigeweg2013, fabricius1998, steinigeweg2009,
steinigeweg2011-1}, $T > 0$ Lanczos methods \cite{prelovsek2013,
mierzejewski2010, steinigeweg2012-1}, quantum Monte-Carlo techniques
\cite{alvarez2002, grossjohann2010}, as well as time-dependent
density matrix renormalization group (tDMRG) approaches to the
real-time dynamics of wave packets or correlators \cite{langer2009,
jesenko2011, karrasch2012, karrasch2013-1, karrasch2013-2,
huang2013, karrasch2014-1, karrasch2014-2} and to the solution of the
Lindblad quantum master equation \cite{prosen2009, znidaric2011}. The
overwhelming majority of results for the spin Drude weight, however, is
only available from ED and tDMRG. Since ED is at present restricted to
chains of length $L \sim 20$, the long-time/low-frequency limit is
still governed by finite-size effects and intricate extrapolation
schemes to the thermodynamic limit have been invoked, with different
results depending on details. Such details are using even or odd $L$
\cite{herbrych2011} and choosing grand-canonical and canonical
ensembles \cite{karrasch2013-1}. Clearly, these details should be
irrelevant in the thermodynamic limit. Alternatively, tDMRG is
exceedingly more powerful w.r.t.\ system size and chains of length
$L \sim 200$ are accessible. But still the method is confined to a
maximum time scale depending on the exchange anisotropy $\Delta$
\cite{karrasch2012, karrasch2013-1, karrasch2014-1, karrasch2014-2}.
Even though there is an ongoing progress to increase this time scale,
it is at present too short for a reliable extraction of the spin Drude
weight at the isotropic point $\Delta = 1$ \cite{karrasch2013-1}, which
is both, experimentally relevant and theoretically most challenging. In
this situation, typicality can provide a fresh numerical perspective
\cite{steinigeweg2014-1}.

Much less is known on transport dynamics apart from the mere
existence of Drude weights. For spin transport at $\Delta = 1$,
steady-state bath scenarios \cite{prosen2009, znidaric2011} and
classical simulations \cite{alcantarabonfim1992, gerling1993,
steinigeweg2012-2} suggest super-diffusive dynamics in the limit of
high temperatures, while bosonization predicts diffusion at low but
nonzero temperatures \cite{sirker2009}. At $\Delta > 1$, signatures
of diffusion have been observed also at high temperatures in
different approaches \cite{prosen2009, znidaric2011,
steinigeweg2012-2, steinigeweg2011-1, steinigeweg2011-2,
steinigeweg2010} (see Ref.\ \onlinecite{karrasch2014-1} for a
discussion of the limit $\Delta \to \infty$). Diffusion of heat,
however, necessarily requires integrability-breaking perturbations.
For nonintegrable problems, perturbation theory is the only
analytical technique available but hard to perform in the
thermodynamic limit \cite{jung2006, jung2007}. The perturbative
regime of small integrability-breaking is further challenging for
numerical methods since dynamics is slow and physically relevant
time scales are long. These long time scales are a challenge for
tDMRG and, due to finite-size effects, also for ED. Thus, typicality
may complement both numerical methods in this demanding regime.

In this paper, we use the concept of typicality to study the
real-time dynamics of spin and energy currents in integrable and
nonintegrable spin-$1/2$ chains at nonzero temperatures. In this
way, we extend our previous work in Ref.\
\onlinecite{steinigeweg2014-2} to energy transport and nonintegrable
systems as well. Within the framework of linear response theory, we
numerically obtain autocorrelation functions from the propagation of
a single pure state, drawn at random as a typical representative of
the full statistical ensemble. By comparing to small-system data
from ED and existing short-time data from tDMRG, we show that
typicality is satisfied in finite systems down to low temperatures
and holds in both, integrable and nonintegrable systems. In
particular, we demonstrate two numerical advantages of typicality:
First, for integrable systems, we can calculate the long-time
dynamics and extract the Drude weight for large systems outside the
range of ED. Thus, typicality improves the reliability of
finite-size scaling. Second, for nonintegrable systems, we can
obtain the full relaxation curve for large systems with little
finite-size effects on the physically relevant time scale. Hence,
typicality provides also the basis for determining the dc
conductivity in the regime of small integrability-breaking model
parameters without using any fits/extrapolations. Both advantages
yield significant progress in the numerical investigation of spin
and energy dynamics in particular and of other observables in a much
broader context.

This paper is structured as follows: In Sec.\ \ref{models} we first
introduce the two models studied, namely, the integrable XXZ
spin-$1/2$ chain and a nonintegrable version due to the presence of
a staggered magnetic field oriented in $z$ direction. In this Sec.
we also define the spin and energy currents, as well as their
time-dependent autocorrelation functions, and we discuss symmetries.
The next Sec.\ \ref{typicality} is devoted to the concept of
dynamical typicality and the closely related numerical technique
used throughout this paper. Then we turn to our results: In Sec.\
\ref{integrable} we focus on spin-current dynamics in the integrable
model and study the spin Drude weight as a function of anisotropy
and temperature. In the following Sec.\ \ref{nonintegrable} we
extend our study in two directions: the nonintegrable model and the
dynamics of the energy current. In this Sec.\ we analyze the
dependence of the dc conductivity on magnetic field, anisotropy, and
temperature. The last Sec.\ \ref{summary} closes with a summary and
draws conclusions.

\section{Models, Currents, and Autocorrelations}
\label{models}

\subsection{Integrable Model}

In this paper we investigate the antiferromagnetic XXZ spin-$1/2$
chain. We employ periodic boundary conditions and write the
well-known Hamiltonian ($\hbar = 1$)
\begin{equation}
H = \sum_{r=1}^L h_r \label{H}
\end{equation}
as a sum over the local energy
\begin{equation}
h_r =  J \, (S_r^x S_{r+1}^x + S_r^y S_{r+1}^y + \Delta \, S_r^z
S_{r+1}^z) \, .
\end{equation}
$S_r^i$, $i=x,y,z$ are the components of spin-$1/2$ operators at
site $r$ and $L$ is the total number of sites. $J > 0$ is the
antiferromagnetic exchange coupling constant and $\Delta$ is the
exchange anisotropy in $z$ direction. For $|\Delta| \leq 1$, the
model in Eq.\ (\ref{H}) has no gap and an antiferromagnetic ground
state; for $|\Delta| > 1$, a gap opens and, for $\Delta < -1$, the
ground state becomes ferromagnetic\cite{kolezhuk2004,
karrasch2013-1}.

In general, Eq.\ (\ref{H}) is integrable in terms of the Bethe
Ansatz \cite{zotos1999, benz2005} and has several symmetries. Two
commonly employed symmetries are the invariance under rotation about
the $z$ axis, i.e., the conservation of $S^z = \sum_r S_r^z$, and
translation invariance. Because of these symmetries, the
longitudinal spin $M = -L/2 + i$, $i = 0, 1, \ldots, L$ and the
momentum $k = 2 \pi i/L$ , $i = 0, 1, \ldots, L-1$ are good quantum
numbers. Therefore, the full Hilbert space with $d = 2^L$ states
consists of $(2L+1) \, L$ uncoupled symmetry subspaces with
\begin{equation}
d_{k,M} \approx \frac{1}{L} \binom{L}{|M|+L/2}
\end{equation}
states, where the largest subspaces have $M = 0$. In this paper we
do not restrict ourselves to a specific choice of $M$ or $k$ and
take into account all $(M,k)$ subspaces.

Since the longitudinal magnetization $S^z$ and energy $H$ are
conserved, their currents are well defined operators and follow
from the continuity equation
\begin{equation}
\dot{\rho}_r = \imath [H, \rho_r] = j_{r-1} - j_r \, ,
\end{equation}
where $\rho_r$ is either the local magnetization, $\rho_r = S_r^z$,
or the local energy, $\rho_r = h_r$, and $j_r$ is the corresponding
local current. In case of magnetization, $j_r = -\imath [S_r^z,
h_r]$ and the total current has the well-known form (see, e.g., the
review in Ref.\ \onlinecite{heidrichmeisner2007})
\begin{equation}
j_\text{S} = J \sum_{r=1}^L (S_r^x S_{r+1}^y - S_r^y S_{r+1}^x) \, ,
\label{JS}
\end{equation}
where the subscript $\text{S}$ indicates spin/magnetization for the
remainder of this paper. This current commutes with the Hamiltonian
only for anisotropy $\Delta = 0$ \cite{heidrichmeisner2007}. In the
case of energy, $j_r = -\imath [h_r, h_{r+1}]$ and the total current
can be written as \cite{steinigeweg2013}
\begin{eqnarray}
j_\text{E} &=& J^2 \sum_{r=1}^L [(S_r^x S_{r+2}^y - S_r^y S_{r+2}^x)
S_{r+1}^z \nonumber \\
&-& \Delta (S_r^x S_{r+1}^y - S_r^y S_{r+1}^x) (S_{r-1}^z +
S_{r+2}^z)] \, , \label{JE}
\end{eqnarray}
where the subscript $\text{E}$ indicates energy now. In contrast to
the operators in Eqs.\ (\ref{H}) and (\ref{JS}), the current in Eq.\
(\ref{JE}) acts on more than two neighboring sites and involves
three adjacent sites. This current and the Hamiltonian commute with
each other for all values of $\Delta$ \cite{zotos1997,
kluemper2002}. Both, energy and spin current share the good quantum
numbers $(M,k)$ of the Hamiltonian.

\subsection{Nonintegrable Model}

To break the integrability of the model, we add to Eq.\ (\ref{H})
the term
\begin{equation}
H_B = B \sum_{r=1}^L (-1)^r S_r^z \, ,
\label{HB}
\end{equation}
where $B$ is the strength of a staggered magnetic field oriented in
$z$ direction. While this term does not change the above symmetries
of the model, we have momentum $k = 2 \pi i/(L/2)$, $i = 0, 1,
\ldots, (L-1)/2$ now and symmetry subspaces are twice as large as
before.

Due to the form of $H_B$, neither the definition of the spin current
in Eq.\ (\ref{JS}) nor the definition of the energy current in Eq.\
(\ref{JE}) change. In general, $j_\text{S}$ is independent of any
spatial profile of the magnetic field. But $j_\text{E}$ does not
change because of the staggered profile of $H_B$ in Eq.\ (\ref{HB}).
A homogenous magnetic field, for instance, yields a magnetothermal
correction to the energy current \cite{heidrichmeisner2007}. Such a
correction does not occur in our case. One important consequence of
adding $H_B$ is a nonvanishing commutator
\begin{equation}
[H + H_B, j_\text{E}] = [H_B, j_\text{E}] \propto B \, \Delta \, ,
\label{commutator}
\end{equation}
i.e., the energy current is not strictly conserved for finite values
of $B$ and $\Delta$. Note that this commutator also leads to
scattering rates $1/\tau \propto (B \, \Delta)^2$, as discussed in
more detail later and in Appendix \ref{perturbation}.

While the main physical motivation of adding a staggered magnetic
field $H_B$ is both, breaking integrability and inducing current
scattering, we do not intend to describe a specific experimental
situation. Moreover, choosing $H_B$ as perturbation allows us to
compare with existing results in the literature.

\subsection{Autocorrelation Functions}

Within the framework of linear response theory \cite{kubo1991}, we
investigate spin and energy autocorrelation functions at inverse
temperatures $\beta = 1/T$ ($k_B = 1$),
\begin{eqnarray}
C_\text{S/E}(t) &=& \frac{\text{Re} \, \langle j_\text{S/E}(t) \,
j_\text{S/E} \rangle}{L} \nonumber \\
&=& \frac{\text{Re} \, \text{Tr} \{e^{-\beta H} j_\text{S/E}(t) \,
j_\text{S/E} \}}{L \, Z} \, , \label{ACF}
\end{eqnarray}
where the time argument of the operator $j_\text{S/E}(t)$ has to be
understood w.r.t.\ the Heisenberg picture, $j_\text{S/E} =
j_\text{S/E}(0)$, and $Z = \text{Tr}\{ \exp(-\beta H) \}$ is the
partition function. In the limit of high temperatures $\beta \to 0$,
the sum rules are given by $C_\text{S}(0) = J^2/8$ and
$C_\text{E}(0) = J^4(1 + 2 \Delta^2)/32$ \cite{zotos1997,
kluemper2002}. Linear response theory describes the dynamics close
to equilibrium and is valid in both, integrable and nonintegrable
systems \cite{mierzejewski2010, steinigeweg2012-1}. For
nonequilibrium effects, see Ref.\ \onlinecite{mierzejewski2010}.

Spin and energy autocorrelation functions in Eq.\ (\ref{ACF}) can be
written as
\begin{equation}
C_\text{S/E}(t) = \sum_{k,M} \sum_{n,n'}^{d_{k,M}} \frac{e^{-\beta
n}}{Z} \frac{|\langle n | j_\text{S/E} | n' \rangle|^2}{L}
\cos(\omega_{nn'} \, t) \, , \label{ED}
\end{equation}
where $n = n(k,M)$ labels eigenstates of the Hamiltonian in the
symmetry subspace $(k,M)$ and $\omega_{nn'} = E_n - E_{n'}$ is the
difference of eigenvalues $E_n$ and $E_{n'}$. The expression in Eq.\
(\ref{ED}) provides the basis for exact-diagonalization studies
\cite{narozhny1998, heidrichmeisner2003, heidrichmeisner2007,
herbrych2011, steinigeweg2013, fabricius1998, steinigeweg2009,
steinigeweg2011-1}. In these studies, $C_\text{S/E}(t)$ can be
easily evaluated for arbitrarily long times; however, eigenstates
and eigenvalues can be obtained only for systems of finite size.
Even with symmetry reduction, accessible sizes are $L \sim 20$ as of
today. For such systems, the long-time behavior of $C_\text{S/E}(t)$
can be affected by strong finite-size effects, also in the limit of
high temperatures $\beta \to 0$. Generally, finite-size effects
increase as temperature is lowered and are stronger for integrable
systems \cite{heidrichmeisner2007}.

We are interested in extracting information on two central transport
quantities, namely, the Drude weights $\bar{C}_\text{S/E}$ and the
{\it regular} dc conductivities $\kappa_\text{S/E}$,
\begin{equation}
\bar{C}_\text{S/E} = \int \limits_{t_1}^{t_2} \! \text{d}t \,
\frac{C_\text{S/E}(t)}{t_2 - t_1} \, , \quad \kappa_\text{S/E} =
\beta \int \limits_0^{t_3} \! \text{d}t \, C_\text{S/E}(t) \, .
\label{TQ}
\end{equation}
Here, $t_2 \gg t_1 \gg 1/J$, with $t_1$ and $t_2$ selected from a
region where $C_\text{S/E}(t)$ has practically decayed to its
long-time value $C_\text{S/E}(t \to \infty) \geq 0$, which may be
zero or nonzero. We emphasize that, with this selection,
$\bar{C}_\text{S/E}$ can be safely viewed as time-independent and is
the Drude weight \cite{zotos1997, kluemper2002, shastry1990,
zotos1999, benz2005, prosen2011, prosen2013, narozhny1998,
heidrichmeisner2003, heidrichmeisner2007, herbrych2011,
steinigeweg2013, fujimoto2003}. A nonzero Drude weight exists
whenever the current is at least partially conserved and therefore
indicates ballistic transport. In cases where the Drude weight
vanishes and transport is not ballistic in the thermodynamic limit,
the dc conductivities $\kappa_\text{S/E} = \kappa_\text{S/E}(\omega
\rightarrow 0^+)$ are of interest and result from a zero-frequency
Fourier transform of $C_\text{S/E}(t)$, i.e., the right expression
in Eq.\ (\ref{TQ}) with $t_3 \rightarrow \infty$. In finite systems,
however, the Drude weight may be tiny but is never zero. Thus, if
the limit $t_3 \to \infty$ is performed in a finite system,
$\kappa_\text{S/E}$ will always diverge. To avoid such divergences
for cases with tiny Drude weights, we choose a finite but long
cutoff time $t_3 \gg \tau$, where $\tau$ is the relaxation time,
i.e., $C_\text{S/E}(\tau) / C_\text{S/E}(0) = 1/e$. For cases with a
clean exponential relaxation, for instance, $t_3 = 5 \tau$ is a
suitable choice because times $> 5 \tau$ do not contribute
significantly to $\kappa_\text{S/E}$. Hence, for these cases,
choosing $t_3 = 5 \tau$ yields a reasonable approximation of the dc
conductivity on the basis of a finite system. In general, one has to
ensure that $\kappa_\text{S/E}$ is approximately independent of the
specific choice of $t_3$.

Note that other definitions of the Drude weight exist in the
literature, where additional prefactors $\pi$, $2 \pi$, or $\beta$
appear. Note further that our expression for the energy conductivity
is different from other definitions using a prefactor $\beta^2$. In
this way, we use the expression in Ref.\ \onlinecite{huang2013}.

\begin{figure}[tb]
\includegraphics[width=0.8\columnwidth]{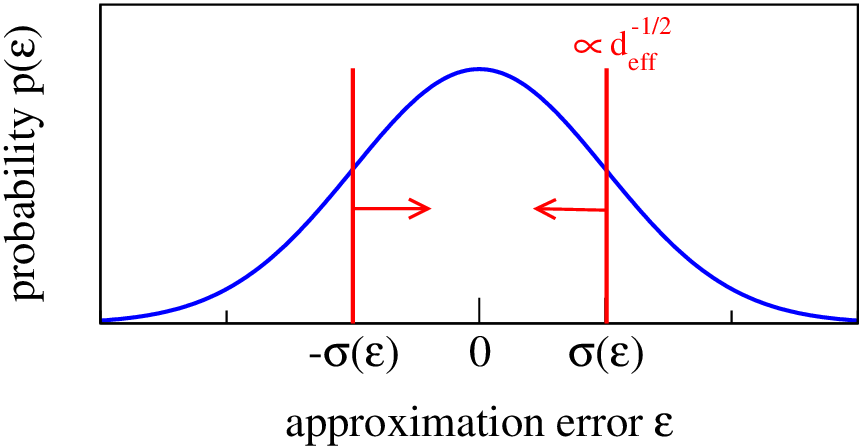}
\caption{(color online) Sketch of the probability distribution
$p(\epsilon)$ for the approximation error $\epsilon(| \psi \rangle)$
in Eq.\ (\ref{approximation1}). While the average of the error
vanishes, $\bar{\epsilon} = 0$, the variance of the error decreases
as the effective Hilbert-space dimension increases,
$\sigma(\epsilon) \propto 1/\sqrt{d_\text{eff}}$. This scaling
implies that $p(\epsilon)$ is a $\delta$ function in the
thermodynamic limit.} \label{Fig1}
\end{figure}

\section{Dynamical Typicality}
\label{typicality}

\subsection{Approximation}

Next we introduce an approximation of autocorrelation functions.
This approximation provides the basis of the numerical method used
throughout this paper. The central idea amounts to replacing the
trace $\text{Tr}\{\bullet\} = \sum_i \langle i | \bullet | i
\rangle$ in Eq.\ (\ref{ACF}) by a single scalar product $\langle
\psi | \bullet | \psi \rangle$ involving a pure state $|\psi
\rangle$, which, furthermore, is drawn at random. Since we aim at
the dynamics in the full Hilbert space, $|\psi \rangle$ is randomly
chosen in the full basis. This is conveniently done in the common
eigenbasis of symmetries,
\begin{equation}
| \psi \rangle = \sum_{M,k} | \psi_{M,k} \rangle \, , \quad
|\psi_{M,k} \rangle = \sum_s^{d_{M,k}} (a_s + \imath \, b_s) \, | s
\rangle \, , \label{PS}
\end{equation}
where $s = s(k,M)$ is a label for the common eigenstates of
symmetries and $a_s$, $b_s$ are random real numbers. Specifically,
$a_s$, $b_s$ are chosen according to a Gaussian distribution with
zero mean. In this way, the pure state $| \psi \rangle$ is chosen
according to a distribution that is invariant under all unitary
transformations in Hilbert space (Haar measure \cite{bartsch2009})
and, according to typicality \cite{gemmer2003, goldstein2006,
reimann2007, popescu2006, white2009, sugiura2012}, a representative
of the statistical ensemble.

$| \psi \rangle$ and $| \psi_{M,k} \rangle$ in Eq.\ (\ref{PS})
correspond to high temperatures $\beta \to 0$. To incorporate
finite temperatures, we introduce $| \psi_{M,k}(\beta) \rangle =
\exp(-\beta H/2) \, | \psi_{M,k} \rangle$. Then, we rewrite the
autocorrelation function in Eq.\ (\ref{ACF}) as \cite{hams2000,
bartsch2009, elsayed2013, steinigeweg2014-1, steinigeweg2014-2}
\begin{eqnarray}
C_\text{S/E}(t) &=& \frac{\text{Re} \, \sum_{M,k} \langle
\psi_{M,k}(\beta) | j_\text{S/E}(t) \, j_\text{S/E} |
\psi_{M,k}(\beta) \rangle}{L \, \sum_{M,k} \langle \psi_{M,k}(\beta)
| \psi_{M,k}(\beta) \rangle} \nonumber
\\
&+& \epsilon_\text{S/E}(| \psi \rangle) \, , \label{approximation1}
\end{eqnarray}
where $\epsilon_\text{S/E}(| \psi \rangle)$ encodes the error which
results if the first term on the r.h.s.\ of Eq.\
(\ref{approximation1}) is taken as an {\it approximation} for
$C_\text{S/E}(t)$. This error is random due to the random choice of
$| \psi \rangle $. Certainly, one may sample over several $| \psi
\rangle$ and in fact $\epsilon_\text{S/E}$ vanishes, i.e.,
$\bar{\epsilon}_\text{S/E} = 0$. This sampling is routinely done to
obtain autocorrelation functions in the context of classical
mechanics \cite{steinigeweg2012-2}.

The main point of Eq.\ (\ref{approximation1}), however, is that, in
addition to the mean error $\bar{\epsilon}_\text{S/E} = 0$, one also
knows the standard deviation of errors
$\sigma(\epsilon_\text{S/E})$, as illustrated in Fig.\ \ref{Fig1}.
Precisely, one knows an upper bound for
$\sigma(\epsilon_\text{S/E})$ \cite{bartsch2009, elsayed2013,
steinigeweg2014-1},
\begin{equation}
\sigma(\epsilon_\text{S/E}) \leq {\cal O} \left (
\frac{\sqrt{\text{Re} \, \langle j_\text{S/E}(t) \, j_\text{S/E} \,
j_\text{S/E}(t) \, j_\text{S/E} \rangle}} {L \, \sqrt{d_\text{eff}}}
\right ) \, , \label{error}
\end{equation}
where $d_\text{eff}$, the effective dimension of the Hilbert space,
occurs. In the high-temperature limit $\beta \to 0$, $d_\text{eff} =
2^L$ is identical to the full dimension $d$. Thus, if the number of
sites $L$ is increased, $\sigma(\epsilon_\text{S/E})$ decreases
exponentially fast with $L$. Therefore, remarkably, the {\it single}
pure-state contribution from the first term on the r.h.s.\ of Eq.\
(\ref{approximation1}) turns into an exponentially good
approximation for the autocorrelation function. At arbitrary
$\beta$, $d_\text{eff} = \text{Tr} \{ \exp[-\beta(H-E_0)] \}$ is the
partition function and $E_0$ the ground-state energy, i.e.,
$d_\text{eff}$ essentially is the number of thermally occupied
states. Again, this number scales exponentially fast with $L$
\cite{steinigeweg2014-1} but less quickly. To summarize, discarding
$\epsilon_\text{S/E}$ in Eq.\ (\ref{approximation1}) is an
approximation which is exact in the thermodynamic limit $L \to
\infty$. At finite $L$, errors can be reduced additionally by
sampling, however, this sampling turns out to be unnecessary for all
examples in this paper.

\begin{figure}[tb]
\includegraphics[width=0.8\columnwidth]{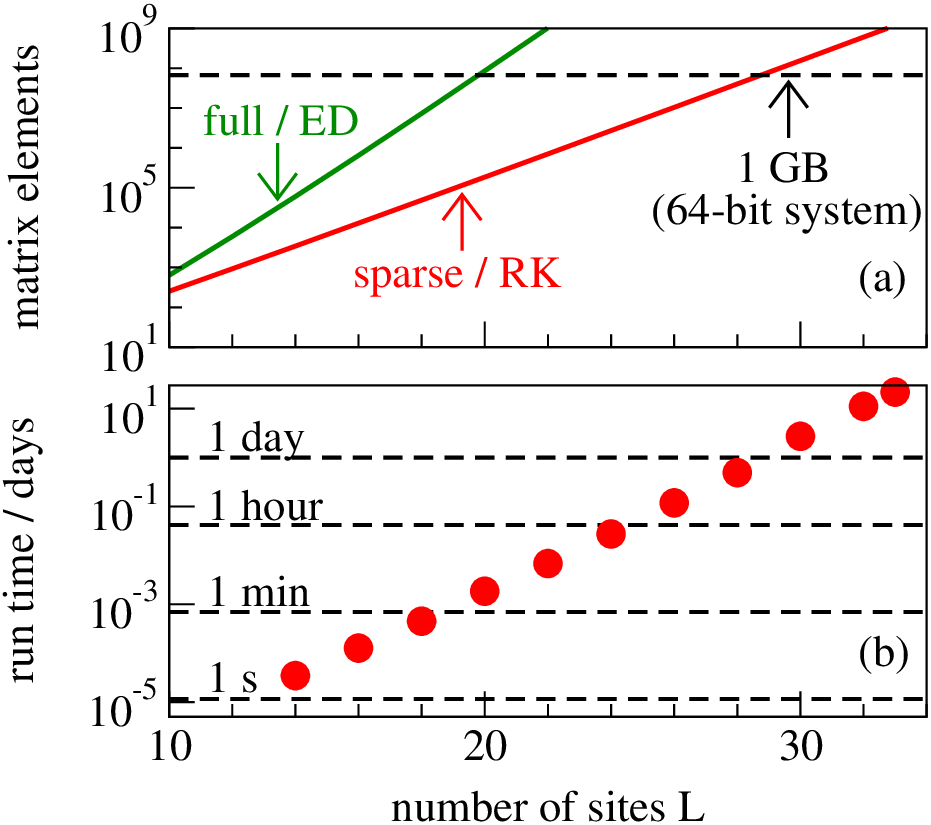}
\caption{(color online) (a) Total number of matrix elements required
to represent the largest subspaces ($M=0$,$k$) of the Hamiltonian in
Eq.\ (\ref{H}) as a full and sparse matrix. The full matrix for
$L=20$ requires the same amount of computer memory as the sparse
matrix for $L=30$. (b) Run time of the Runge-Kutta algorithm for the
same subspaces and the spin current using a single CPU and $10^4$
discrete time steps ($\delta t \, J = 0.01$, $t \, J = 100$). The
run time for $L=33$ is roughly one month. (Note that adding Eq.\
(\ref{HB}) to the Hamiltonian increases  matrix elements and run
time by a factor $2$.)} \label{Fig2}
\end{figure}

\subsection{Numerical Method}

The central advantage of the approximation in Eq.\
(\ref{approximation1}) is that it can be calculated without the
eigenstates and eigenvalues of the Hamiltonian, in contrast to the
exact expression in Eq.\ (\ref{ED}). Specifically, this calculation
is based on the two auxiliary pure states
\begin{eqnarray}
&& | \Phi_{M,k}(\beta,t) \rangle = e^{-\imath H t -\beta H/2} \, |
\psi_{M,k} \rangle \, ,  \label{state1} \\
&& | \varphi_{M,k}(\beta,t) \rangle = e^{-\imath H t} \,
j_\text{S/E} \, e^{-\beta H/2} \, |\psi_{M,k} \rangle \, ,
\label{state2}
\end{eqnarray}
both depending on time and temperature. Note that the only
difference between the two states is the additional current
operator $j_\text{S/E}$ in the r.h.s.\ of Eq.\ (\ref{state2}). By
the use of these states, we can rewrite the approximation in Eq.\
(\ref{approximation1}) as
\begin{equation}
C_\text{S/E}(t) = \frac{\text{Re} \, \sum_{M,k} \langle
\Phi_{M,k}(\beta,t) | j_\text{S/E} | \varphi_{M,k}(\beta,t)
\rangle}{L \, \sum_{M,k} \langle \Phi_{M,k}(\beta,0) |
\Phi_{M,k}(\beta,0) \rangle} \, , \label{approximation2}
\end{equation}
where we skip the error $\epsilon_\text{S/E}(| \psi \rangle)$ for
clarity. Apparently, a time dependence of the current operator
$j_\text{S/E}$ does not appear anymore in Eq.\
(\ref{approximation2}). Instead, the full time and temperature
dependence is a property of the pure states only.

The $\beta$ dependence, e.g.\ of $| \Phi_{M,k}(\beta,t)\rangle$,
is generated by an imaginary-time Schr\"odinger equation,
\begin{equation}
\imath \, \frac{\partial}{\partial (\imath \beta)} \, |
\Phi_{M,k}(\beta,0) \rangle = \frac{H}{2} \,  | \Phi_{M,k}(\beta,0)
\rangle \, , \label{imagS}
\end{equation}
and the $t$ dependence by the usual real-time Schr\"odinger
equation,
\begin{equation}
\imath \, \frac{\partial}{\partial t} \, | \Phi_{M,k}(\beta,t)
\rangle = H \,  | \Phi_{M,k}(\beta,t) \rangle \, . \label{realS}
\end{equation}
These differential equations can be solved by the use of
straightforward iterator methods, e.g.\ Runge-Kutta
\cite{elsayed2013, steinigeweg2014-1, steinigeweg2014-2}, or more
sophisticated Chebyshev \cite{deraedt2007, jin2010} schemes. In this
paper, we use a fourth order Runge-Kutta (RK4) scheme with a
discrete time step $\delta t \, J = 0.01 \ll 1$. For this small
$\delta t$, numerical errors are negligible, as shown later by the
time-independent norm of $| \Phi_{M,k}(\beta,t) \rangle$ and $|
\varphi_{M,k}(\beta,t) \rangle$ and the agreement with results from
other methods.

In the Runge-Kutta scheme, we have to implement the action of
Hamiltonian and currents on pure states. It is possible to carry out
these matrix-vector multiplications without saving matrices in
computer memory. Therefore, the memory requirement of the algorithm
is set only by the size of vectors: ${\cal O}(d_{M,k})$. However, to
reduce the run time of the algorithm, it is convenient to save
matrices in memory. In this respect, we can profit from the fact
that Hamiltonian and currents are few-body operators with a
sparse-matrix representation, even in the common eigenbasis of
symmetries. In fact, for all operators, there are only $L \, d_{M,k}
\ll d_{M,k}^2$ nonvanishing matrix elements, as illustrated in Fig.\
\ref{Fig2} (a). Thus, the memory requirement of the algorithm is
${\cal O}(L \, d_{M,k})$ and scales linearly with the dimension of
the symmetry subspaces. Consequently, we are able to treat chains
with as many as $L=33$ sites, where the largest subspaces at $M=0$
are huge:
\begin{equation}
d_{0,k} \approx 3.5 \cdot 10^7 \, .
\end{equation}
As compared to the upper subspace dimension accessible to exact
diagonalization, this dimension is orders of magnitude larger, i.e.,
by a factor ${\cal O}(10^4)$. Note that the dimension of the full
Hilbert space is $d \approx 10^{10}$. In Fig.\ \ref{Fig2} (b) we
show the run time of the algorithm for the largest subspaces and the
spin current using a single CPU and $10^4$ discrete time steps
($\delta t \, J = 0.01$, $t \, J = 100$). For $L= 33$, the run time
is about one month while, for $L=20$, the calculation takes about
two minutes. We note that, for cases where less time steps are
needed ($\delta t \, J = 0.01$, $t \, J \ll 100$), $L>33$
calculations are also feasible within reasonable run time, as
demonstrated in Sec.\ \ref{L34}.

Obviously, iterator methods for solving Eqs.\ (\ref{imagS}) and
(\ref{realS}) are very typical approaches to profit from massive
parallelization. This has been pursued in other applications of
typicality, {\it neglecting} however the impact of symmetry
reduction \cite{deraedt2007}. Regarding our algorithm, the latter
adds an additional layer of paralellization, i.e., due to the good
quantum numbers $(M,k)$, each of the $(2L + 1) L$ subspaces can be
computed independently. Remarkably, in practice, we have only relied
on the latter paralellization, using not more than $L$ CPUs on
medium-sized clusters. In this way, we were able to reach system
sizes identical to those of massively parallelized codes on super
computers without symmetry reduction \cite{deraedt2007, jin2010}. We
believe that symmetry reduction in combination with massive
parallelization has the potential to reach $L \sim 40$ in the
future.

\section{Spin-Current Dynamics in the Integrable Model}
\label{integrable}

\begin{figure}[tb]
\includegraphics[width=1.0\columnwidth]{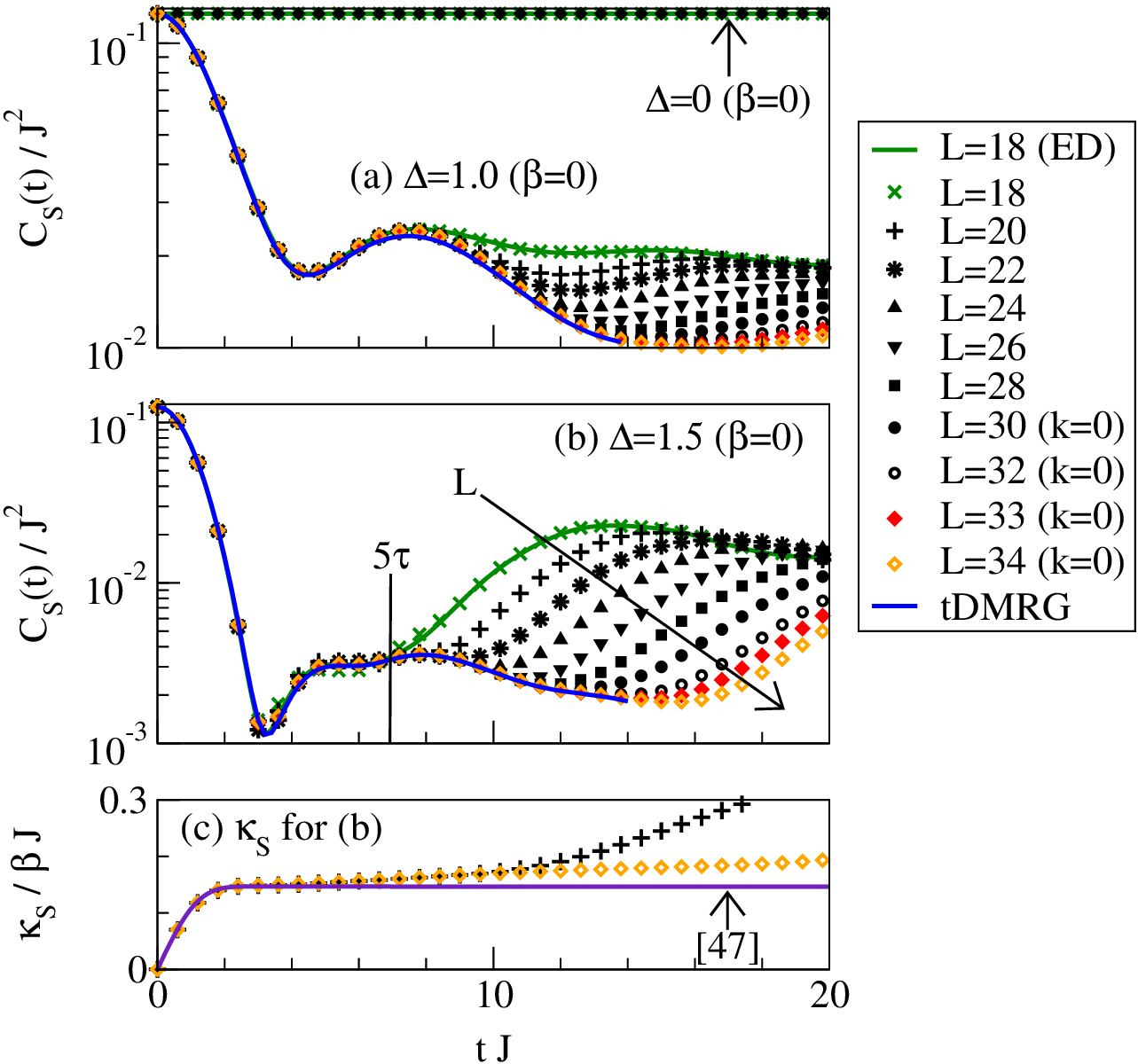}
\caption{(color online) Spin-current autocorrelation function
$C_\text{S}(t)$ at $\beta \to 0$ for (a) $\Delta = 1.0$ and $0$, (b)
$\Delta = 1.5$, numerically obtained for $L =18$ using the exact
expression in Eq.\ (\ref{ED}) (green curve) and larger $L \geq 18$
using the approximation in Eq.\ (\ref{approximation2}) (symbols),
shown in a semi-log plot. The very high accuracy of the
approximation is illustrated by comparing to available tDMRG data
for $L=200$ \cite{karrasch2012, karrasch2014-1} (blue curve). (c) Dc
conductivity $\kappa_\text{S}$ for (b), evaluated according to Eq.\
(\ref{TQ}) as a function of the cutoff time. For comparison, the
perturbative result of Ref.\ \onlinecite{steinigeweg2011-1} is
depicted (curve).} \label{Fig3}
\end{figure}

\begin{figure}[t]
\includegraphics[width=0.8\columnwidth]{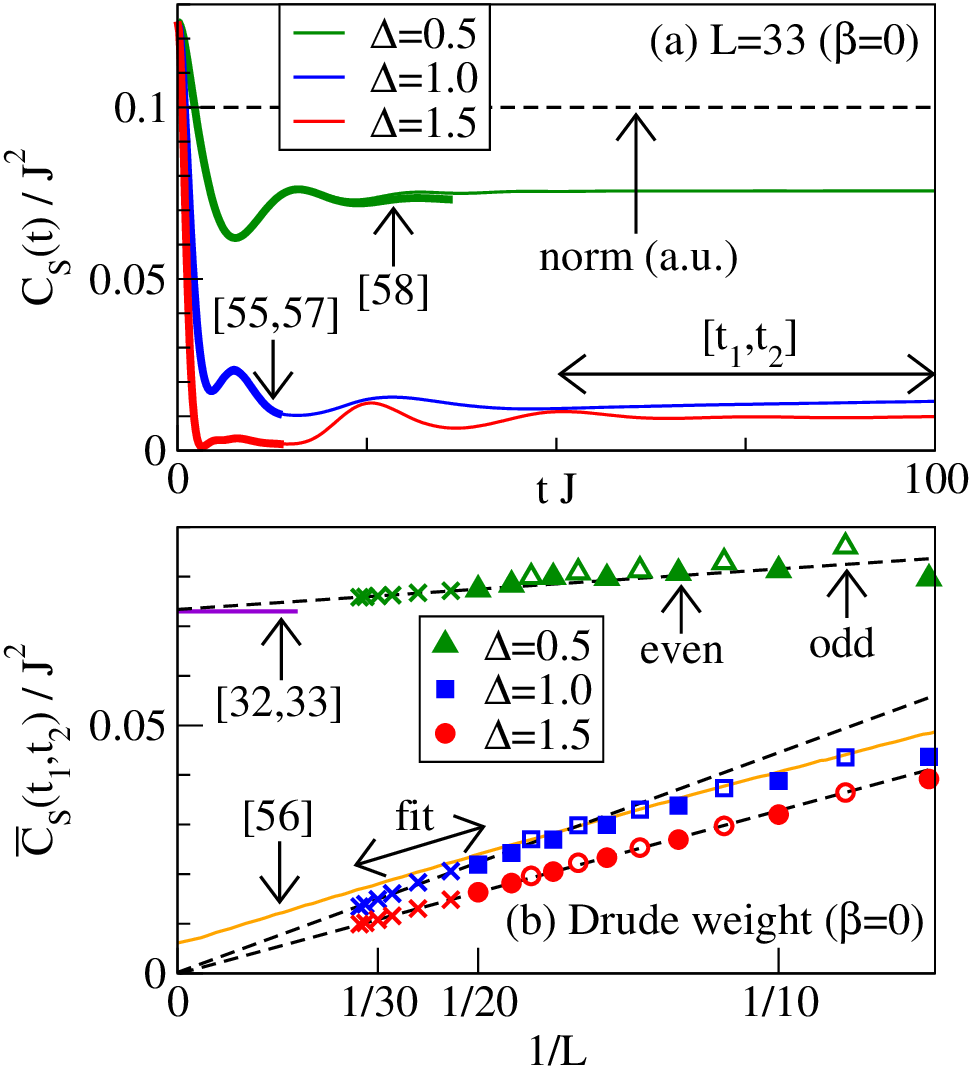}
\caption{(color online) (a) Long-time limit of the spin-current
autocorrelation function $C_\text{S}(t)$ at $\beta \to 0$ for
$\Delta = 0.5$, $1.0$, and $1.5$, numerically obtained using the
approximation in Eq.\ (\ref{approximation2}) (thin solid curves).
The well-conserved norm is indicated (thin dashed curve). Available
tDMRG data for $L=200$ \cite{karrasch2012, karrasch2014-1,
karrasch2014-2} is depicted (thick solid curves). (b) Finite-size
scaling of the spin Drude weight, extracted according to Eq.\
(\ref{TQ}) in the time interval $[t_1 \, J, t_2 \, J] = [50,100]$
(small $L \leq 20$: exact expression in Eq.\ (\ref{ED}), closed
(open) symbols for even (odd) $L$; large $L > 20$: approximation in
Eq.\ (\ref{approximation2}), crosses). Simple $1/L$ fits to large
$20 \leq L \leq 33$ are shown (dashed lines), and at $\Delta = 1.0$
the odd-site fit to $L \leq 19$ performed in
Ref.~\onlinecite{karrasch2013-1} as well (solid curve). At small
$\Delta = 0.5$, T.\ Prosen's rigorous analytic lower bound
\cite{prosen2011, prosen2013} is indicated (horizontal line).}
\label{Fig4}
\end{figure}

\begin{figure}[b]
\includegraphics[width=0.8\columnwidth]{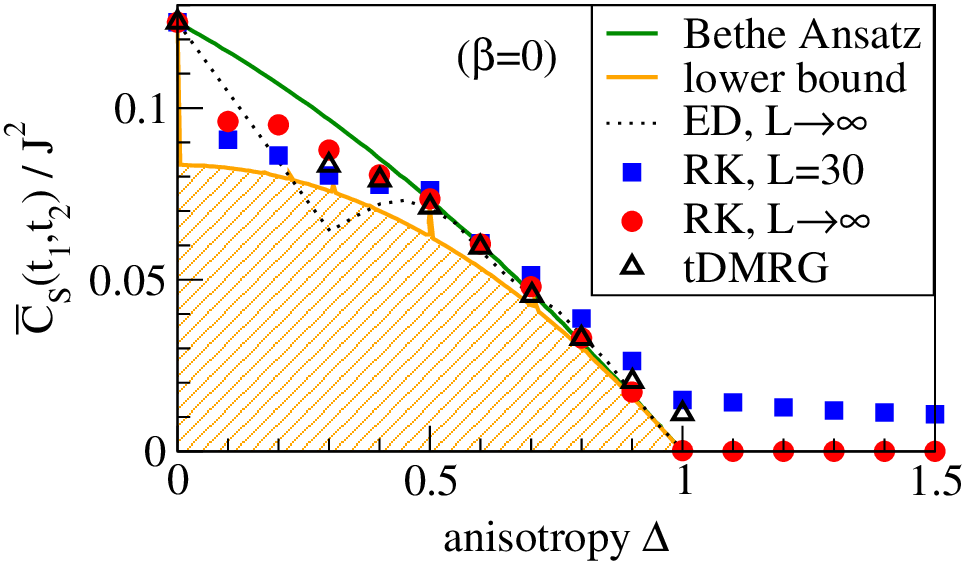}
\caption{(color online) High-temperature spin Drude weight
$\bar{C}_\text{S}$ w.r.t.\ the anisotropy $\Delta$, obtained from
the approximation in Eq.\ (\ref{approximation2}) using Runge-Kutta
(closed symbols). Results are compared to the thermodynamic Bethe
Ansatz \cite{zotos1999}, T.\ Prosen's strict analytic lower bound
\cite{prosen2011, prosen2013}, the extrapolation based on exact
diagonalization at zero magnetization and odd sites in Ref.\
\onlinecite{herbrych2011}, and tDMRG \cite{karrasch2012} (see also
Ref.\ \onlinecite{karrasch2013-1} for a different point of view on
the tDMRG data point at $\Delta = 1$).} \label{Fig5}
\end{figure}

First, we present results on the integrable model in Eq.\ (\ref{H}).
Since the energy current $j_\text{E}$ is strictly conserved in this
model, we focus on the dynamics of the spin current $j_\text{S}$.
Parts of the corresponding results in Figs.\ \ref{Fig3}, \ref{Fig4},
and \ref{Fig5} have been shown in our previous work
\onlinecite{steinigeweg2014-2}. Subsequently, we also investigate
the dynamics of the energy current for the nonintegrable model.

\subsection{High Temperatures and Intermediate Times}
\label{L34}

We begin with the high-temperature limit $\beta \to 0$ and
intermediate times $t \, J \leq 20$. For anisotropy $\Delta = 1$ and
small $L=18$, we compare in Fig.\ \ref{Fig3} (a) the exact and
approximate expressions of the autocorrelation function in Eqs.\
(\ref{ED}) and (\ref{approximation2}), numerically calculated by the
use of exact diagonalization and fourth order Runge-Kutta,
respectively. For all times, the agreement between Eqs.\ (\ref{ED})
and (\ref{approximation2}) is  remarkably good. Our usage of a
semi-log plot underlines this agreement even more and emphasizes
relative rather than only absolute accuracy. Due to this agreement,
we can already consider the approximation as almost exact for
$L=18$. Moreover, any remaining error decreases exponentially fast
with $L$. Thus, we can safely neglect any averaging over random pure
states $| \psi \rangle$. Note that, for the models studied in this
paper, significant errors only occur below $L \sim 10$, see Appendix
\ref{small}.

By increasing $L$ in Fig.\ \ref{Fig3} (a), we show that the curve of
the autocorrelation function gradually converges in time towards the
thermodynamic limit. For the maximum size $L=34$ calculated, the
curve is converged up to times $t \, J \sim 15$ with no visible
finite-size effects in the semi-log plot. For the four largest $L
\geq 30$ depicted, we restrict ourselves to a single translation
subspace, i.e., $k = 0$, to reduce computational effort in the
high-temperature limit $\beta \to 0$. For these temperatures, it is
already known that the $k$ dependence is negligibly small
\cite{herbrych2011}, and we also do not observe a significant
dependence on $k$ for $L < 30$, see Appendix \ref{independence}.

Additionally, we compare to existing tDMRG data for a system of very
large size $L = 200$ \cite{karrasch2012}. It is intriguing to see
that our results agree up to very high precession. On the one hand,
this very good agreement is a convincing demonstration of dynamical
typicality in an integrable system. On the other hand, this
agreement unveils that our numerical technique yields exact
information on an extended time window in the thermodynamic limit $L
\to \infty$. As shown later, this time window can become very large
for nonintegrable systems.

In Fig.\ \ref{Fig3} (b) we show a second calculation for a larger
anisotropy $\Delta = 1.5$. Clearly, the autocorrelation function
decays to almost zero rapidly. However, there is a small long-time
tail. This tail has been observed already on the basis of exact
diagonalization for intermediate $L=20$ \cite{steinigeweg2009}. It
is not connected to the Drude weight \cite{karrasch2014-1}, as
discussed in more detail later. It is also not a finite-size effect,
as evident from the agreement with tDMRG. While this tails shows the
tendency to decay when comparing $t J \sim 10$ and $15$, we relate
its origin to the onset of revivals in the vicinity of the Ising
limit \cite{karrasch2014-1}. If we partially neglect the tail and
determine the dc conductivity $\kappa_S$ according to Eq.\
(\ref{TQ}) for $t_3 \, J = 5 \tau \, J = 7$, we get $\kappa_S/(\beta
J) = 0.159$. This value agrees well with the theoretically predicted
value $0.147$ of the perturbation theory in Ref.\
\onlinecite{steinigeweg2011-1}, see Fig.\ \ref{Fig3} (c). Because
this theory does not take into account the tail, both values
slightly differ. For the maximum time without finite-size effects,
$t_3 \, J = 10 \tau \, J = 14$, we get $\kappa_S/(\beta J) = 0.178$.
This value is still not larger than $20\%$ of theoretical
prediction, see also Ref.\ \onlinecite{karrasch2014-1}.

\subsection{Long-Time Limit}

Next, we investigate the long-time limit. In Fig.\ \ref{Fig4} (a) we
show $C_\text{S}(t)$ in the high-temperature limit $\beta \to 0$ and
various anisotropies $\Delta = 0.5$, $1.0$, and $1.5$. Additionally,
we depict the norm of $| \Phi_\beta(t) \rangle$. This norm is
practically constant, as the norm of $| \varphi_\beta(t) \rangle$
also. The constant norm clearly demonstrates that the Runge-Kutta
scheme works properly at such long times. The data in Fig.\
\ref{Fig4} (a) also proves the saturation of $C_\text{S}(t)$ at
rather long time scales $t J \sim 50$. Furthermore, we can hardly
infer the saturation value from our short-time data in Fig.\
\ref{Fig3}. We note that no fluctuations are visible in the
long-time limit since the approximation error of our numerical
approach is exponentially small and practically zero for the large
system sizes depicted.

In contrast to short times, the long-time limit is still governed by
finite-size effects. Hence, we are now going to perform a proper
finite-size scaling for the Drude weight. We use the definition of
the Drude weight according to Eq.\ (\ref{TQ}) and average over the
time interval $[t_1 \, J, t_2 \, J] = [50,100]$, without invoking
assumptions. The Drude weight has been extracted the same way in
Ref.~\onlinecite{steinigeweg2009}. Moreover, this way of extracting
the Drude weight reproduces the correct zero-frequency values for
small $L$ in, e.g., Ref.\ \onlinecite{heidrichmeisner2003}.

In Fig.\ \ref{Fig4} (b) we depict the resulting Drude weight vs.\
the inverse length $1/L$ for anisotropies $\Delta = 0.5$, $1.0$, and
$1.5$. For $L > 20$, we extract the Drude weight from the
approximation in Eq.\ (\ref{approximation2}) (denoted by crosses)
and, for $L \leq 20$, we use the exact expression in Eq.\ (\ref{ED})
(denoted by other symbols). In this way, we avoid typicality errors
at small $L$. We also indicate the results of $1/L$ fits, solely
based on data points for $L \geq 20$. In this way, we avoid the need
of $(1/L)^{i>1}$ corrections as well as the influence of even-odd
effects at small $L$ and, especially, at small $\Delta$
\cite{evenodd}, see Fig.\ \ref{Fig4} (b). For the small
$\Delta=0.5$, the resulting fit is close to all data points.
Furthermore, extracting the thermodynamic limit $L \to \infty$ from
the fit, we find a non-zero Drude weight in convincing agreement
with the rigorous lower bound of Refs.~\onlinecite{prosen2011,
prosen2013}. While the situation is rather similar for the large
anisotropy $\Delta = 1.5$, the Drude weight vanishes, in agreement
with previous work \cite{heidrichmeisner2003}. The isotropic point
$\Delta = 1.0$ is certainly the most interesting case. Here, the $L
\geq 20$ fit is not close to the one obtained from only small $L <
20$. In fact, the extrapolation yields much smaller values for the
Drude weight than the finite values suggested in previous works,
based on either smaller $L$ \cite{heidrichmeisner2003,
karrasch2013-1} or shorter $t$ \cite{karrasch2012} (see also Ref.\
\onlinecite{karrasch2013-1} for a comprehensive discussion).
Moreover, our result points to a vanishing Drude weight for $L \to
\infty$.

In Fig.\ \ref{Fig5} we summarize the finite-size values for the
Drude weight for fixed $L=30$ and various anisotropies $0 \leq
\Delta \leq 1.5$. Additionally, we indicate the extrapolated values
for $L \to \infty$ using fits. Since even-odd effects are stronger
closer to the point $\Delta = 0$, we take into account only even
sites for the fits. Remarkably, all extrapolated values lie above
the rigorous lower bound of Refs.\ \onlinecite{prosen2011,
prosen2013}, and in the anisotropy range $0.4 \lesssim \Delta \leq
1.5$, also agree with the Bethe-Ansatz solution of Ref.\
\onlinecite{zotos1999}. They further agree with an alternative
extrapolation on the basis of small $L$ \cite{herbrych2011}, using a
different statistical ensemble and only odd sites. In the vicinity
of the point $\Delta = 0$, we still lie above the lower bound but we
observe deviations from the Bethe-Ansatz result. These deviations do
{\it not} indicate the breakdown of typicality and are well-known to
occur in numerical studies using finite systems \cite{herbrych2011},
due to the very high degeneracy at $\Delta = 0$. As visible in Fig.\
\ref{Fig5}, tDMRG results for $\Delta \sim 0.3$ also show these
deviations and are in very good agreement with our results.

\begin{figure}[tb]
\includegraphics[width=1.0\columnwidth]{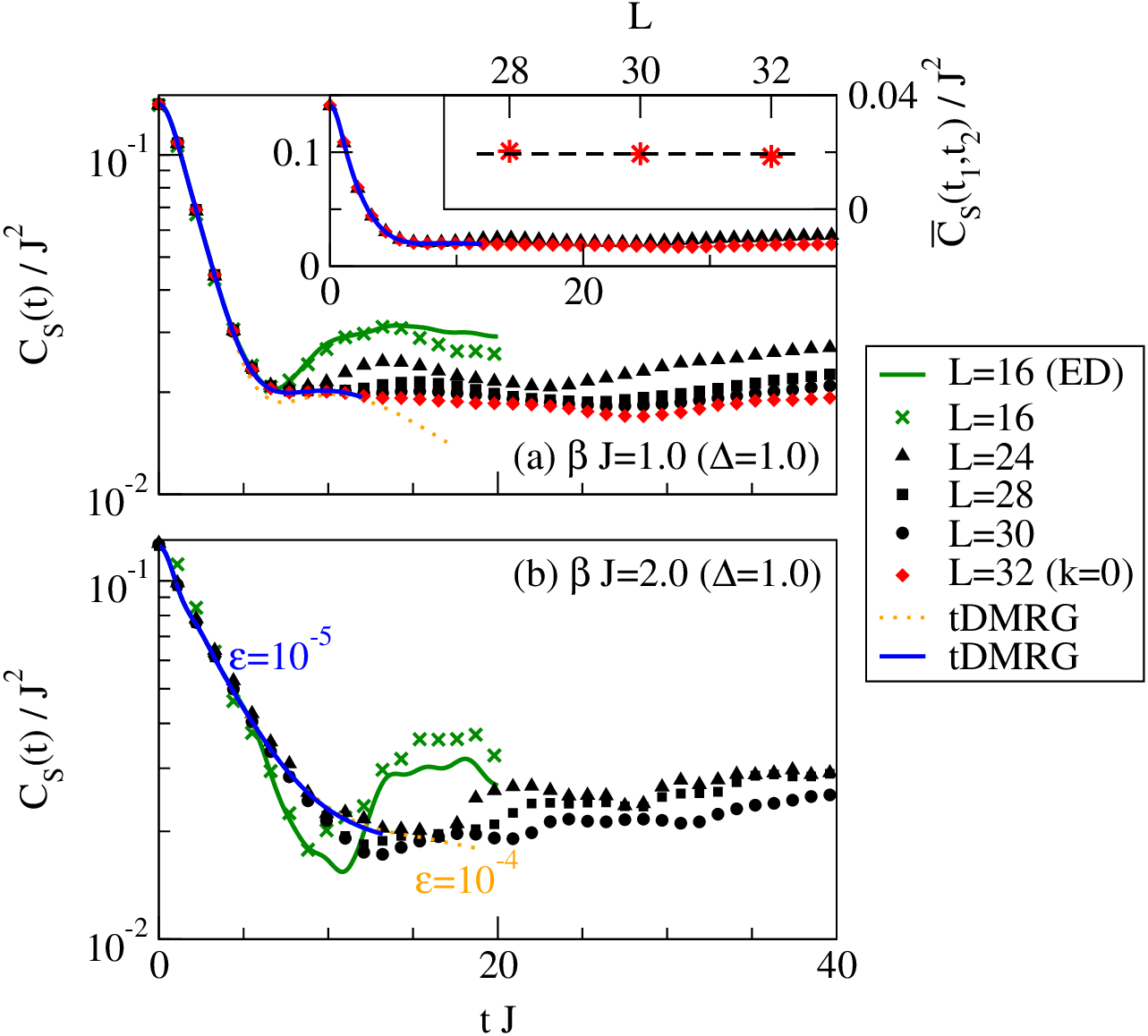}
\caption{(color online) Spin-current autocorrelation function
$C_\text{S}(t)$ for the isotropic point $\Delta = 1.0$ at (a) $\beta
\, J = 1.0$ and (b) $\beta \, J = 2.0$, numerically obtained for $L
=16$ using the exact expression in Eq.\ (\ref{ED}) (green curve) and
larger $L \geq 16$ using the approximation in Eq.\
(\ref{approximation2}) (symbols), shown in a semi-log plot.
Available tDMRG data for $L=200$ \cite{karrasch2012, karrasch2013-1}
is depicted for two different values of the discarded weight
$\epsilon$ (blue and orange curve). Inset in (a): lin-lin plot of
(a) for $L=24$ and $32$; and inset therein: average height
$\bar{C}(t_1,t_2)$ of the plateau at times $[t_1 J,t_2 J] = [10,40]$
for the three largest $L$ where this plateau is clearly seen. The
horizontal line is a guide to the eye.} \label{Fig6}
\end{figure}

\subsection{Low Temperatures}

We now turn to finite temperatures $\beta \neq 0$. Clearly, the
approximation in Eq.\ (\ref{approximation2}) has to break down for
$\beta \to \infty$, i.e., $T \to 0$, due to the reduction of the
effective Hilbert space dimension $d_\text{eff}$. Recall that
$d_\text{eff}$ essentially counts the number of thermally occupied
states. Furthermore, for $\beta \, J \gg 2$, also the exact
expression in Eq.\ (\ref{ED}) is governed by large finite-size
effects, at least for a finite system of size $L \sim 30$
\cite{prelovsek2013}. Thus, for a numerical approach to $L \sim 30$,
reasonable temperatures are $\beta \, J \sim 2$. For this range of
$\beta$, the approximation is still justified and averaging over
pure states is necessary for $\beta \gg 2$ only. This temperature,
however, depends on the specific model, as discussed later in more
detail for the nonintegrable system.

For a small size $L=16$, anisotropy $\Delta = 1$, and the two lower
temperatures $\beta \, J = 1$ and $2$, we compare in Figs.\
\ref{Fig6} (a) and (b) the exact expression in Eq.\ (\ref{ED}) and
the approximation in Eq.\ (\ref{approximation2}), calculated by the
use of exact diagonalization and Runge-Kutta, respectively. Clearly,
deviations appear at $\beta \, J = 2$. However, these deviations
manifest as random fluctuations rather than systematic drifts and
may be compensated by additional averaging over several pure states
$| \psi \rangle$. Furthermore, one can expect that these deviations
disappear for significantly larger sizes $L$. Again, we prove this
expectation by comparing with available tDMRG data for $L=200$
\cite{karrasch2012, karrasch2013-1}. The very good agreement
illustrates the power of our numerical approach at finite
temperatures. Moreover, taking into account the simple structure of
the curve, the semi-log plot, and the combination of tDMRG with our
numerical approach, Fig.\ \ref{Fig6} (a) points to non-zero Drude
weights at $\beta \neq 0$. This observation is different from our
previous results at $\beta = 0$. Still there are finite-size effects
for the three largest values of $L$ where a plateau is clearly seen
at times $[t_1 J,t_2 J] = [10,40]$. But these finite-size effects
are hardly visible in a lin-lin plot, see the inset in Fig.\
\ref{Fig6} (a). Moreover, extrapolations are impossible on the basis
of three and approximately constant points, see the inset therein.

It is further worth mentioning that the possibility of non-vanishing
Drude weights at $\beta \neq 0$ is consistent with the recent upper
bound in Ref.\ \onlinecite{carmelo2014}. This upper bound does {\it not}
vanish when taking into account all sectors of magnetization, as done
in our paper.

\section{Energy-Current Dynamics in the Nonintegrable Model}
\label{nonintegrable}

Next, we extend our analysis in two directions. First, we break the
integrability of the model by the staggered magnetic field in Eq.\
(\ref{HB}). Second, we expand our analysis to include the dynamics
of the energy current $j_\text{E}$, which is {\it not} conserved
anymore in the nonintegrable model.

\begin{figure}[t]
\includegraphics[width=1.0\columnwidth]{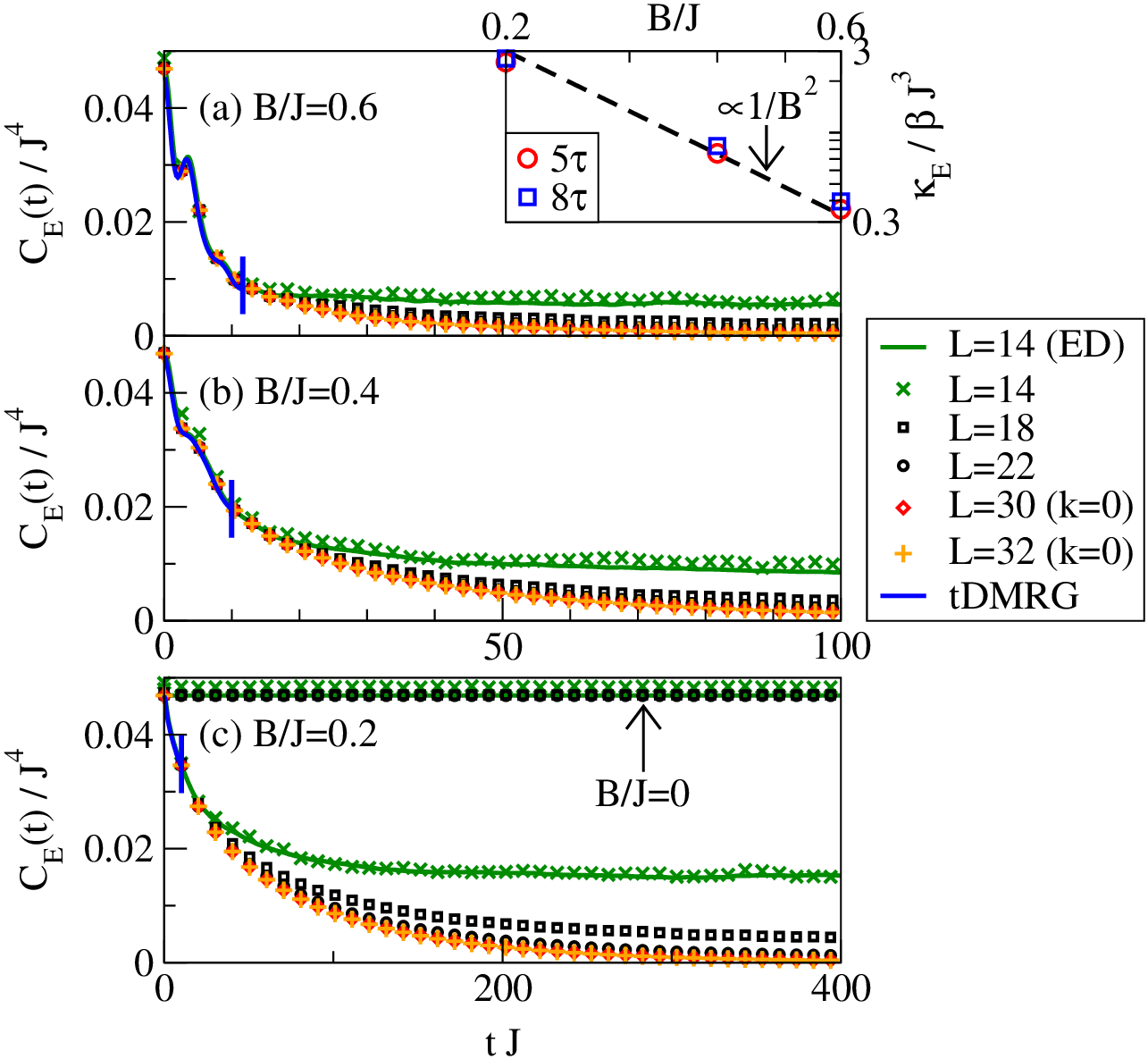}
\caption{(color online) Real-time decay of the energy-current
autocorrelation function $C_\text{E}(t)$ for the XXZ spin-$1/2$
chain at anisotropy $\Delta = 0.5$ in a staggered magnetic field of
strength (a) $B/J = 0.6$, (b) $B/J = 0.4$, and (c) $B/J = 0.2$ in
the high-temperature limit $\beta \, J \to 0$. Numerical results
according to Eq.\ (\ref{approximation2}) for length $L = 14$ agree
well with both, exact diagonalization and available tDMRG data for
$L=200$ in Ref.\ \onlinecite{karrasch2013-2}. (The maximum time of
this tDMRG data is indicated by vertical blue bars.) For $L > 14$,
little finite-size effects are visible up to full relaxation, e.g.,
$L=22$ and $L=32$ are indistinguishable in (a)-(c). Inset in (a):
The dc conductivity $\kappa_\text{E}$ according to Eq.\ (\ref{TQ})
with $t_3 = 5\tau$ and $8 \tau$ scales $\propto 1/B^2$ for small
values of $B$.} \label{Fig7}
\end{figure}

\begin{figure}[b]
\includegraphics[width=0.8\columnwidth]{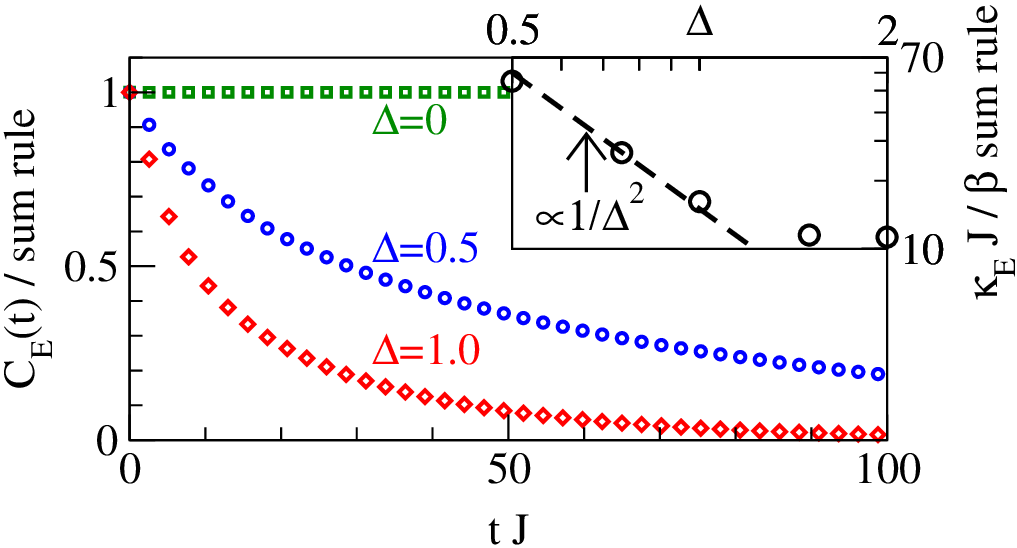}
\caption{(color online) Real-time decay of the energy-current
autocorrelation function $C_\text{E}(t)$ for the XXZ spin-$1/2$
chain at various anisotropies $\Delta = 0$, $0.5$, and $1$ in a
staggered magnetic field of strength $B/J = 0.2$ in the
high-temperature limit $\beta J \to 0$. All initial values are
normalized to $1$, i.e., divided by the sum rule $J^4 (1+2
\Delta^2)/32$. Inset: In units of the sum rule, the dc conductivity
according to Eq.\ (\ref{TQ}) scales $\propto 1/\Delta^2$ for small
values of $\Delta$. The system size is $L=30$ ($k=0$) in all cases.}
\label{Fig8}
\end{figure}

\subsection{Dependence on Magnetic Field and Anisotropy}

Again, we begin with the high-temperature limit $\beta \to 0$ and
compare the exact expression in Eq.\ (\ref{ED}), evaluated by exact
diagonalization, and the approximation in Eq.\
(\ref{approximation2}), evaluated by Runge-Kutta, for the energy
current $j_\text{E}$. We show this comparison in Figs.\ \ref{Fig7}
(a)-(c) for  a small size $L = 14$, anisotropy $\Delta = 0.5$, and
magnetic fields of different strength $B/J = 0.6$, $0.4$, and $0.2$.
Apparently, Eqs.\  (\ref{ED}) and (\ref{approximation2}) agree well
with each other. This good agreement proofs that dynamical
typicality is neither a particular property of the spin current
$j_\text{S}$ nor restricted to the integrable system. Interestingly,
our $L = 14$ data already reproduces existing tDMRG data for $L=200$
in Ref.\ \onlinecite{karrasch2013-2}. We note that
exact-diagonalization data for $L = 12$ does so also, although not
shown here explicitly.

For small $L = 14$ and all $B/J \neq 0$ in Fig.\ \ref{Fig7} (a)-(c),
the energy-current autocorrelation function $C_\text{E}(t)$ does not
decay to zero and features a nonzero Drude weight. The actual value
of the finite-size Drude weight increases as $B$ is decreased.
However, by increasing $L$, we show that $C_\text{E}(t)$ decays to
zero for significantly larger $L$ and all $B$ considered. Moreover,
we find that $C_\text{E}(t)$ is practically the same for $L=22$ and
$L=32$. This finding indicates little finite-size effects up to full
relaxation. Hence, our numerical approach yields exact information
on the full, physically relevant time window in the thermodynamic
limit $L \to \infty$.

Let us discuss the relaxation curve for large $L$ in more detail.
The relaxation time decreases as $B$ increases and the overall
structure of the curve is simple, in particular without any slowly
decaying long-time tails. Therefore, extracting from our numerical
data the dc conductivity $\kappa_\text{E}$ according to Eq.\
(\ref{TQ}) yields similar values for cutoff times $t_3 = 5 \tau$ and
$8 \tau$. These values are shown in the inset of Fig.\ \ref{Fig7}
(a). The apparent decrease of $\kappa_\text{E}$ with $B$ results
from the decrease of $\tau$ with $B$ and our usage of a log-log plot
unveils the scaling $\kappa_\text{E} \propto 1/B^2$, as expected
from conventional perturbation theory at small $B$
\cite{steinigeweg2010, steinigeweg2011-2}. Note that the commutator
in Eq.\ (\ref{commutator}) essentially is the memory kernel of the
perturbation theory and yields scattering rates $1/\tau \propto (B
\, \Delta)^2/(1 + 2 \Delta^2)$ and hence, in units of the sum rule,
the scaling
\begin{equation}
\kappa_\text{E}' = \frac{\kappa_\text{E}}{1 + 2 \Delta^2} \propto
\frac{1 + 2 \Delta^2}{(B \, \Delta)^2} \, ,
\end{equation}
i.e., $\kappa_\text{E} = \kappa_\text{E}' \propto 1/(B \, \Delta)^2$
for small values of $\Delta$. A more detailed description of the
perturbation theory is given in Appendix \ref{perturbation}.

To verify the $\Delta$ dependence from perturbation theory, we
calculate in Fig.\ \ref{Fig8} the energy-current autocorrelation
function $C_\text{E}(t)$ for different $\Delta = 0$, $0.5$ and $1$,
at fixed magnetic field $B/J = 0.2$ and size $L = 30$. Indeed,
$C_\text{E}(t)$ decays the slower the smaller $\Delta$ and no
dynamics occurs at $\Delta = 0$. Moreover, extracting the dc
conductivity $\kappa_\text{E}'$ from our numerical data, we find the
scaling $\kappa_\text{E}' \propto 1/\Delta^2$ at small $\Delta$, as
shown in the inset of Fig.\ \ref{Fig8}. This scaling turns into
$\kappa_\text{E}' = \text{const.}$ at large $\Delta$, still
consistent with the expectation from perturbation theory.

\begin{figure}[t]
\includegraphics[width=1.0\columnwidth]{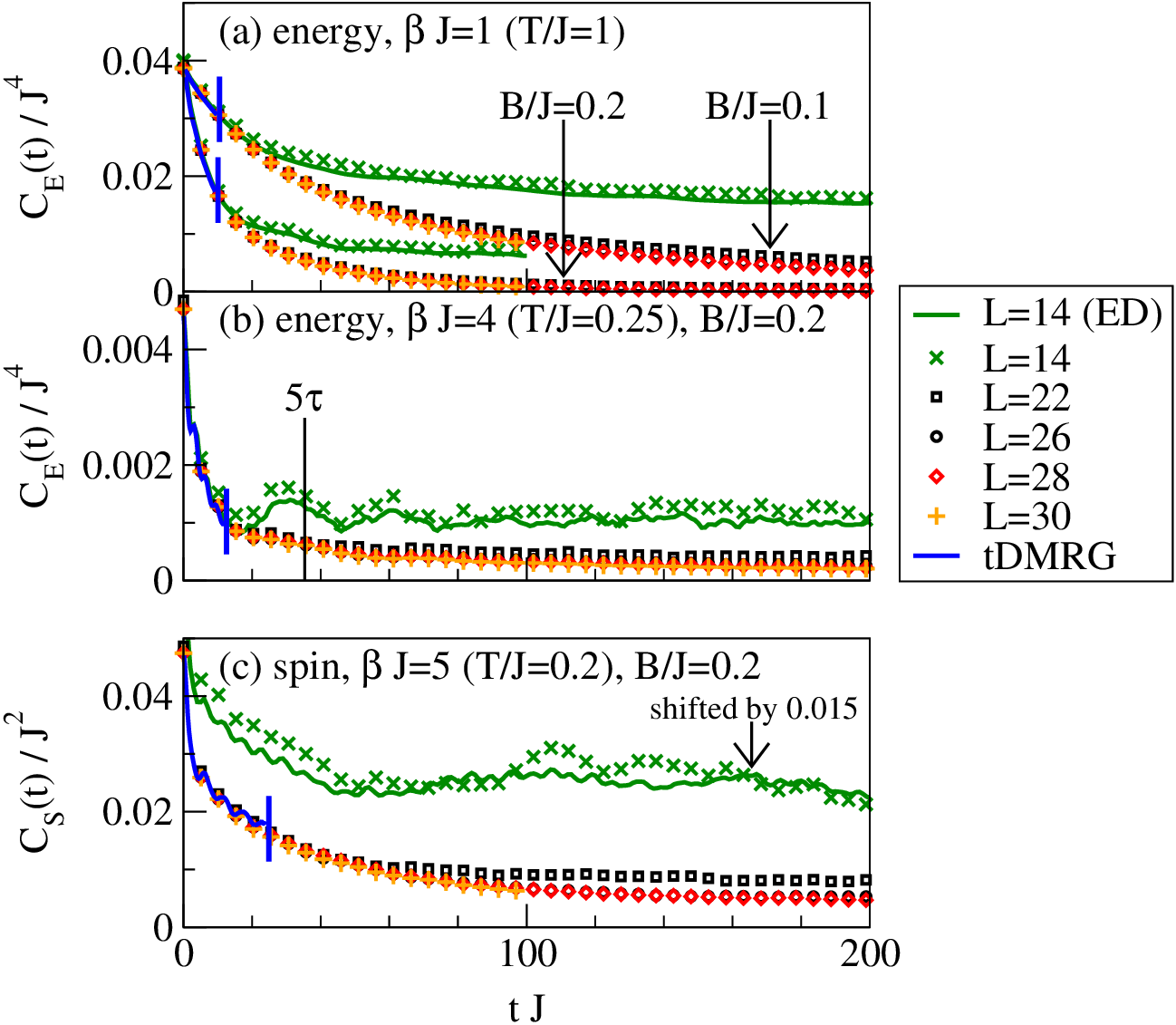}
\caption{(color online) Real-time decay of the autocorrelation
functions of (a), (b) energy current and (c) spin current for the
Heisenberg spin-$1/2$ chain at anisotropy $\Delta=-0.85$ in a
staggered magnetic field of strength $B/J \leq 0.2$. Numerical
results according to Eq.\ (\ref{approximation2}) for length $L=14$,
$\ldots$, $30$ agree well with available tDMRG data in Ref.\
\onlinecite{huang2013} and show little finite-size effects over the
temperature range $\beta \, J \leq 5$ ($T/J \geq 0.2$). In (c)
$L=14$ data is shifted by $0.015$ since this data is very close to
$L > 14$ data.} \label{Fig9}
\end{figure}

\begin{figure}[b]
\includegraphics[width=0.75\columnwidth]{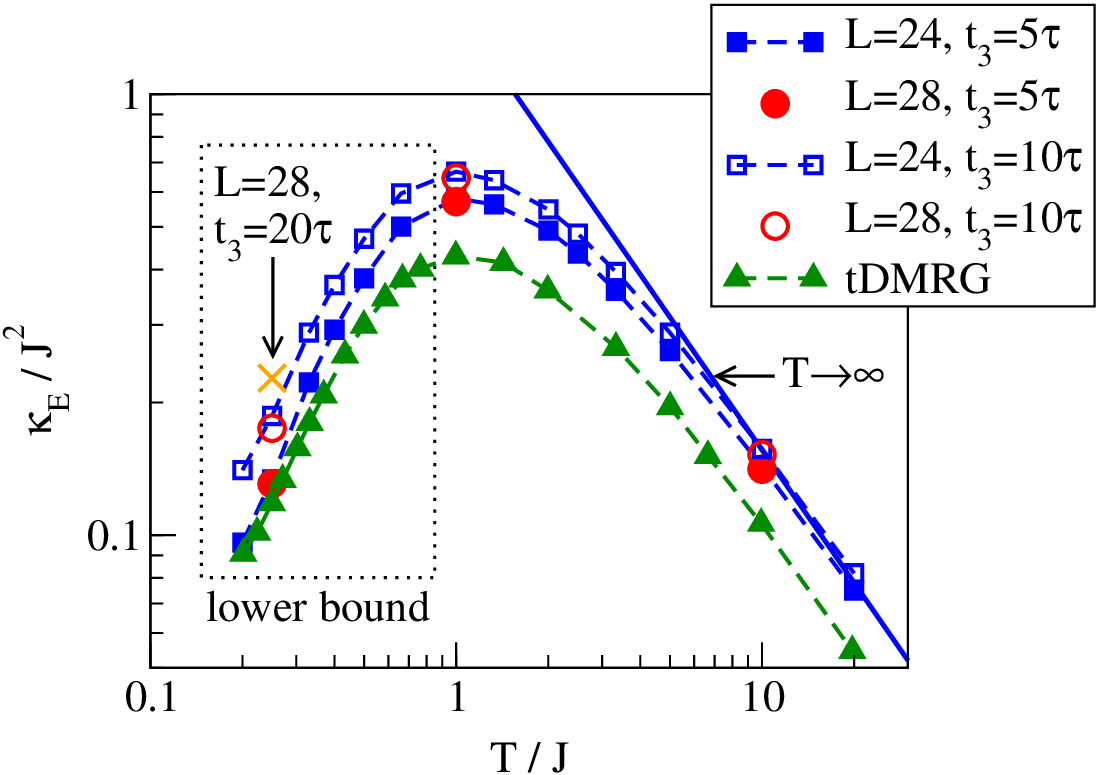}
\caption{(color online) $T$ dependence of the dc conductivity
$\kappa_\text{E}$ for the XXZ spin-$1/2$ chain at anisotropy $\Delta
= -0.95$ in a staggered magnetic field of strength $B/J = 0.2$,
calculated according to Eq.\ (\ref{TQ}) for $t_3 = 5 \tau$  and $10
\tau$. For comparison, results from tDMRG-based fits/extrapolations
in Ref.\ \onlinecite{huang2013} are shown. While the overall
agreement is quite good, deviations are clearly visible and neither
a finite-size effect nor an effect arising from the specific choice
of $t_3$. The dotted box indicates the $T$ region where our results
on $\kappa_\text{E}$ have to be understood as an lower bound.}
\label{Fig10}
\end{figure}

\subsection{Temperature Dependence}

We now turn to finite temperatures $\beta \neq 0$ and choose the
parameters of the model according to the availability of tDMRG data
in the literature. Such data is available for negative anisotropy
$\Delta = -0.85$ \cite{huang2013}, where the model is still
antiferromagnetic. We note that the sign of $\Delta$ has not been of
importance so far since, at high temperatures $\beta \to 0$, the
dynamics depends on $|\Delta|$ only.

In Figs.\ \ref{Fig9} (a) and (b) we summarize our results on the
energy-current autocorrelation function $C_\text{E}(t)$ for an
intermediate temperature $\beta \, J = 1$ and a low temperature
$\beta \, J = 4$. Moreover, we depict our results on the
spin-current autocorrelation function $C_\text{S}(t)$ for an even
lower temperature of $\beta \, J = 5$ in Fig.\ \ref{Fig9} (c). While
the focus is on a magnetic field of strength $B / J = 0.2$, Fig.\
\ref{Fig9} (a) also shows results for the case $B/J = 0.1$. Several
comments are in order. First, already for a small system size of
$L=14$, the exact expression in Eq.\ (\ref{ED}) and the
approximation in Eq.\ (\ref{approximation2}) are in good agreement
at $\beta \, J = 1$. While deviations occur at $\beta \gg 1$, these
deviations are surprisingly small for both, the energy and spin
current in Figs.\ \ref{Fig9} (b) and (c). Second, our
exact-diagonalization data for $L = 14$ already reproduces existing
tDMRG data for $L=200$ in Ref.\ \onlinecite{huang2013} for the whole
temperature range $\beta \, J \leq 5$. This observation is indeed
interesting, especially since lower temperatures have not been
analyzed by tDMRG \cite{huang2013}, at least for the energy current.
Third, our results for large $L$ do not depend significantly on $L$.
This independence demonstrates the high accuracy of the
approximation in Eq.\ (\ref{approximation2}) for large $L$ and also
indicates little (or weakly scaling) finite-size effects.

Because we do not need to deal with finite-size effects, we may
directly extract from our numerical data the dc conductivity
$\kappa_\text{E}$ in Eq.\ (\ref{TQ}), starting with the cutoff time
$t_3 = 5\tau$. This choice has been sufficient for high temperatures
and its role for low temperatures is discussed later in detail. For
the energy current and another negative anisotropy $\Delta = -0.95$,
we show in Fig.\ \ref{Fig10} the resulting temperature dependence of
$\kappa_\text{E}$ in a log-log plot. Apparently, $\kappa_\text{E}
\propto \beta = 1/T$ in the limit of high temperatures $T/J \gg 1$.
This scaling with $T$ is a direct consequence of the trivial
prefactor $\beta$ in Eq.\ (\ref{TQ}) and shows that the actual
energy-current autocorrelation $C_\text{E}(t)$ turns $T$ independent
in that limit. At $T/J \sim 1$, the high-temperature limit is
clearly left and $\kappa_\text{E}(T)$ features a broad maximum. At
$T/J \ll 1$, the low-temperature regime sets in and the scaling of
$\kappa_\text{E}$ with $T$ is consistent with a power law, however,
the exponent remains an open issue. Remarkably, this power-law
scaling mainly results from the initial value $C_\text{E}(0)$ and
not from the time dependence $C_\text{E}(t)$ as such, cf.\ Figs.\
\ref{Fig9} (a) and (b).

In Fig.\ \ref{Fig10} we additionally compare these results on
$\kappa_\text{E}(T)$ with results from tDMRG. Precisely, we compare
to results from fits/extrapolations performed in Ref.\
\onlinecite{huang2013} on the basis of short-time tDMRG data, cf.\
Figs.\ \ref{Fig9} (a) and (b). While the overall agreement is quite
good, our $\kappa_\text{E}(T)$ lies above the one of Ref.\
\onlinecite{huang2013} for all temperatures $0.2 \leq T/J \leq 20$.
We emphasize that this deviation is not a finite-size effect and,
moreover, that it does not result from our choice $t_3 = 5 \tau$ for
the calculation of $\kappa_\text{E}$. In fact, using a longer time
$t_3 = 10 \tau$ yields a positive correction to $\kappa_\text{E}$
and hence increases the deviation, as illustrated in Fig.\
\ref{Fig10}. This correction is again small at high temperatures: At
$T/J = 10$, the correction is only $8\%$ and, at $T/J = 1$, the
correction is slightly higher with $13\%$. But the trend indicates
that $C_\text{E}(t)$ exhibits slowly decaying long-time tails at low
temperatures. Furthermore, such a tail is clearly visible at $T/J =
0.25$ in Fig.\ \ref{Fig9} (b). For this low temperature, our choice
of $t_3 = 5 \tau$ or $10 \tau$ seems to underestimate
$\kappa_\text{E}$ significantly and has to be understood as a {\it
lower bound}. We explicitly avoid analyzing longer $t_3$ since we
cannot exclude the possibility of (weakly scaling) finite-size
effects in the long-time limit.

\begin{figure}[tb]
\includegraphics[width=0.7\columnwidth]{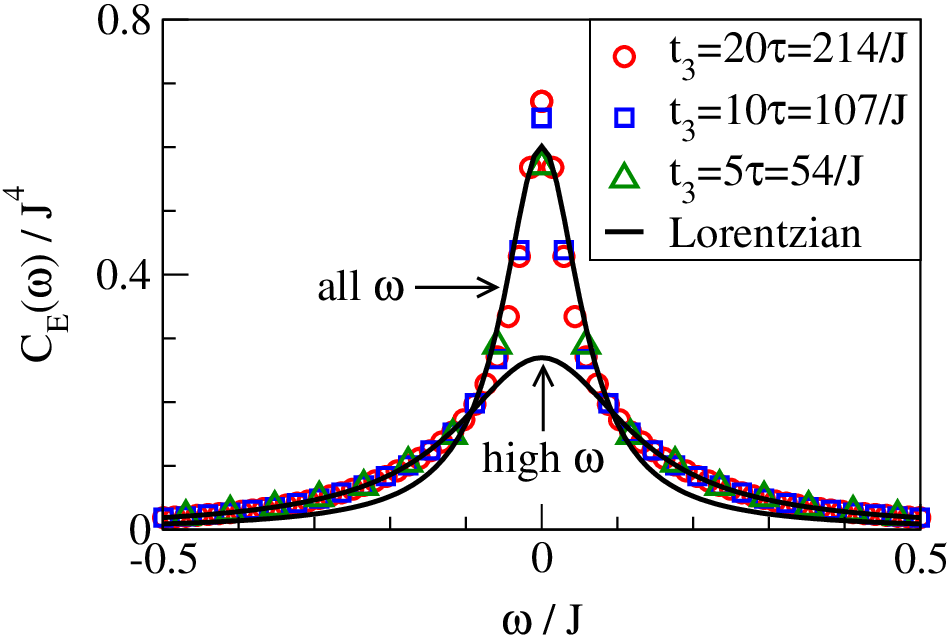}
\caption{(color online) Fourier-transformed energy-current
autocorrelation function $C_\text{E}(\omega)$ for the Heisenberg
spin-$1/2$ chain at anisotropy $\Delta=-0.95$ in a staggered
magnetic field of strength $B/J \leq 0.2$ at temperature $\beta \, J
= 1$. The Fourier transform is performed over time intervals $t_3 =
5 \tau$, $10 \tau$, and $20 \tau$ for a system of size $L=28$. The
results of Lorentzian fits to $C_\text{E}(\omega)$ at all $\omega$
and $|\omega|/J \geq 0.1$ are indicated. } \label{Fig11}
\end{figure}

To gain further insight and to provide an alternative point of view,
we show in Fig.\ \ref{Fig11} the Fourier-transformed energy-current
autocorrelation function $C_\text{E}(\omega)$, still for anisotropy
$\Delta -0.95$ and magnetic field $B/J = 0.2$, at temperature $\beta
\, J = 1$. This parameter set corresponds to the maximum visible in
Fig.\ \ref{Fig10}. The Fourier transform is performed for different
$t_3 = 5 \tau$, $10 \tau$, and $20 \tau$. Note that the longest time
is $20 \tau = 214/J$. While $C_\text{E}(\omega)$ at $|\omega|/J \geq
0.1$ does not depend on the specific choice of $t_3$, it does at
$|\omega|/J \ll 0.1$. Clearly, a minimum time $\sim 10 \tau$ is
required to determine the limit $\omega \to 0$ with sufficient
accuracy, as a consequence of slowly decaying long-time tails at low
temperatures. We also indicate the result of a Lorentzian fit to
$C_\text{E}(\omega)$ at $|\omega|/J \geq 0.1$. Evidently, this
high-frequency/short-time fit cannot be used to predict the dc value
correctly. This fact illustrates the origin of the underestimation
in Ref.\ \onlinecite{huang2013}. We stress that the overall form of
$C_\text{E}(\omega)$ is not Lorentzian at all, while our previous
results in the limit of high temperatures agree well with a
Lorentzian line shape, see Appendix \ref{Fourier}.

\section{Summary}
\label{summary}

In summary, we used the concept of typicality to study the real-time
relaxation of spin and energy currents in spin-$1/2$ chains at
finite temperatures. These chains were the integrable XXZ chain and
a nonintegrable version due to the presence of a staggered magnetic
field oriented in $z$ direction. In the framework of linear response
theory, we numerically calculated autocorrelation functions by
propagating a single pure state, drawn at random as a typical
representative of the full statistical ensemble. By comparing to
data from exact diagonalization for small system sizes and existing
data from tDMRG for short times, we showed that typicality holds in
finite systems over a wide range of temperature and is fulfilled in
both, integrable and nonintegrable systems.

For the integrable model, we calculated the dynamics of the spin
current for long times and extracted the spin Drude weight for large
system sizes outside the range of state-of-the-art exact
diagonalization. Employing proper finite-size scaling, we provided
strong evidence that, at high temperatures above the exchange
coupling constant $J$, the Drude weight vanishes at the isotropic
point. This finding, and also our results for other values of the
exchange anisotropy, were in good agreement with existing
Bethe-Ansatz and Mazur-inequality results. For lower temperatures on
the order of $J$, we found at least indications that the Drude
weight is nonzero at the isotropic point.

For the nonintegrable model, we calculated the decay of the energy
current for large system sizes and did not observe significant
finite-size effects. Therefore, we were able to obtain the full
decay curve in the thermodynamic limit and to extract the dc
conductivity without invoking difficult fits/extrapolations.
Analyzing the dependence of the dc conductivity on the parameters of
the model, we found a quadratic scaling with the inverse magnetic
field and exchange anisotropy, in agreement with conventional
perturbation theory. Moreover, we detailed the temperature
dependence of the dc conductivity, including low- and
high-temperature power laws with an intermediate maximum. Our
numerical results seem to provide a lower bound on the dc
conductivity.

From a merely numerical point of view, we profit from two central
advantages of the typicality-based technique used in this paper.
First, the numerical technique allows us to perform finite-size
scaling. This is certainly similar to exact diagonalization.
However, system sizes are much larger and extrapolations are more
reliable. Second, the numerical technique also yields exact
information on an extended time window in the thermodynamic limit.
This is certainly similar to tDMRG. However, time windows accessible
seem to be much longer for generic nonintegrable systems, as evident
for the example studied in this paper. Because of these two
advantages, our numerical method may complement other numerical
approaches in a much broader context, including problems with few
symmetries and/or in two dimensions. Moreover, our numerical method
may be applied to other observables \cite{elsayed2013} and not only
to current operators.

\section*{Acknowledgments}

We sincerely thank H.\ Niemeyer, P.\ Prelov\v{s}ek, and J.\ Herbrych
for fruitful discussions as well as C.\ Karrasch and F.
Heidrich-Meisner for the tDMRG data in Sec.\ \ref{integrable}
(Refs.\ \onlinecite{karrasch2012, karrasch2013-1, karrasch2014-1,
karrasch2014-2}) and helpful comments. The tDMRG data in Sec.\
\ref{nonintegrable} (Refs.\ \onlinecite{karrasch2013-2, huang2013})
have been digitized.

Part of this work has been done at the Platform for
Superconductivity and Magnetism, Dresden. Part of this work has been
supported by DFG FOR912 Grant No.\ BR 1084/6-2, by SFB 1143, as well
as by EU MC-ITN LOTHERM Grant No.\ PITN-GA-2009-238475.

\appendix

\begin{figure}[t]
\includegraphics[width=0.8\columnwidth]{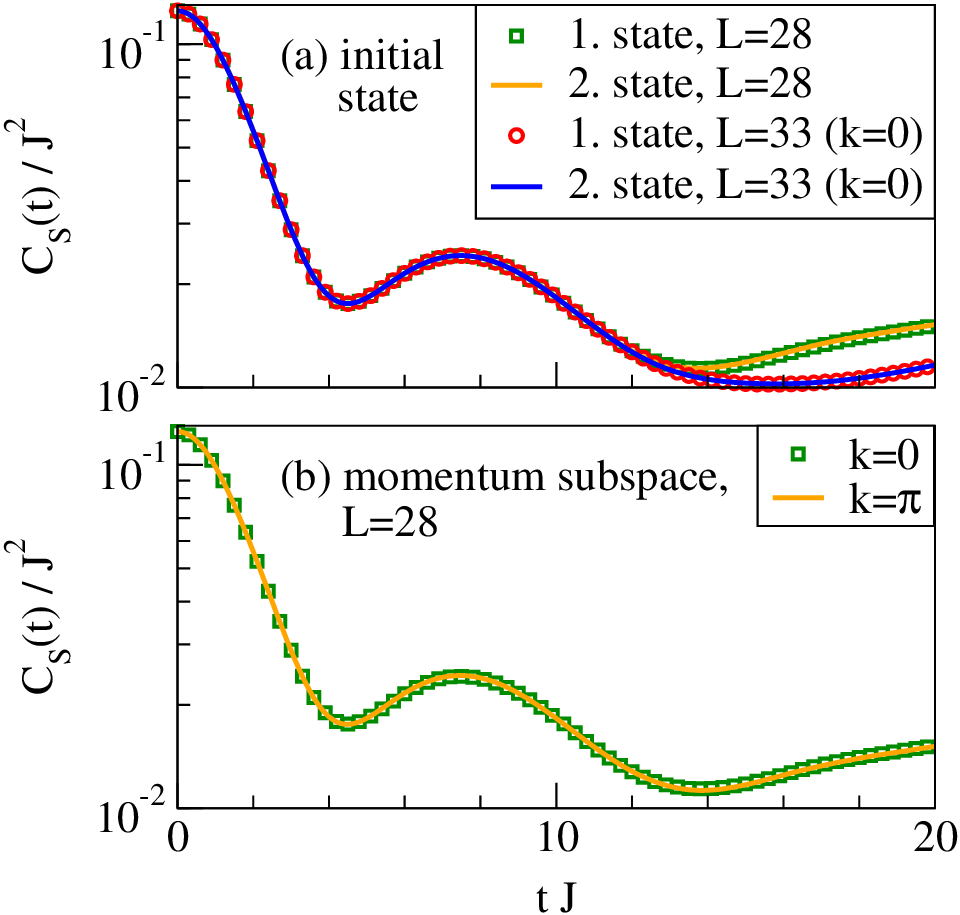}
\caption{(color online) Spin-current autocorrelation function
$C_\text{S}(t)$ in the integrable model at anisotropy $\Delta = 1.0$
and high temperatures $\beta \to 0$ for (a) two different initial
states and (b) two different momentum subspaces, numerically
obtained for $L = 28$ and $33$ using the approximation in Eq.\
(\ref{approximation2}).} \label{Fig12}
\end{figure}

\section{Independence of the Specific Initial State and Momentum
Subspace}
\label{independence}

In Fig.\ \ref{Fig12} we demonstrate that for large system sizes $L$
the approximation in Eq.\ (\ref{approximation2}) depends neither on
the specific realization of the random initial state $| \psi
\rangle$ nor on the momentum subspace $k$ considered. Because both
facts are particularly relevant for the finite-size scaling of the
spin Drude weight in Fig.\ \ref{Fig4}, we present in Fig.\
\ref{Fig12} results for the integrable model in Eq.\ (\ref{H}) at
anisotropy $\Delta = 1$ in the limit of high temperatures $\beta \to
0$. In this limit, the independence of initial states and momentum
subspace holds for all examples given in this paper. For low
temperatures, we explicitly avoid the restriction to a single
momentum subspace since the independence of $k$ is not ensured in
this temperature regime, also for the exact expression in Eq.\
(\ref{ED}).

\begin{figure}[b]
\includegraphics[width=0.7\columnwidth]{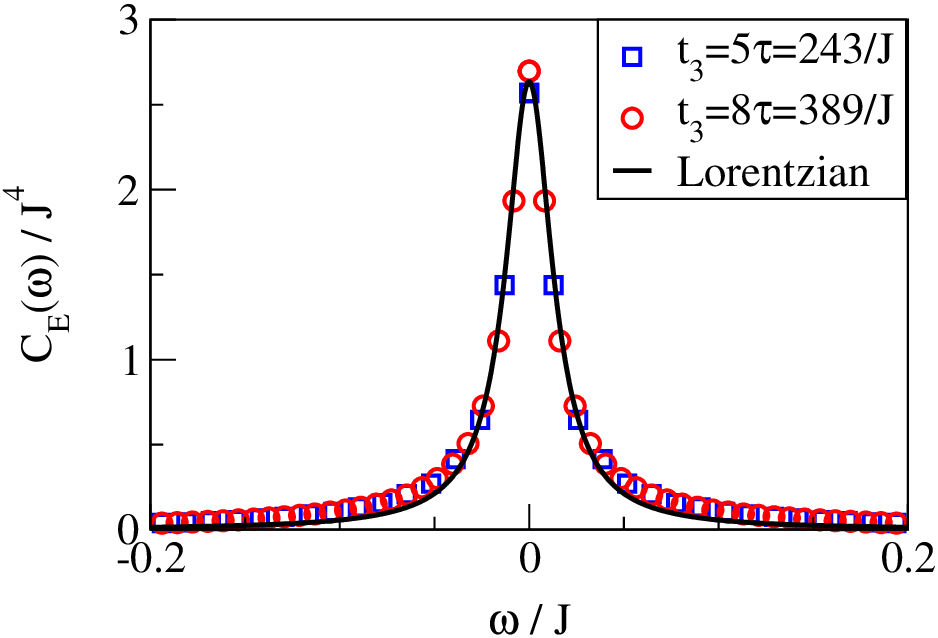}
\caption{(color online) Fourier-transformed energy-current
autocorrelation function $C_\text{E}(\omega)$ for the XXZ spin-$1/2$
chain at anisotropy $\Delta=0.5$ in a staggered magnetic field of
strength $B/J \leq 0.2$ in the high-temperature limit $\beta \to 0$.
The Fourier transform is performed over time intervals $t_3 = 5
\tau$ and $8 \tau$ for system size $L=32$ ($k=0$). The result of a
Lorentzian fit to $C_\text{E}(\omega)$ at all $\omega$ is
indicated.} \label{Fig13}
\end{figure}

\section{Perturbation Theory for the Energy Current}
\label{perturbation}

We discuss here the perturbation theory for the energy current in
detail. Since the energy current $j_\text{E}$ is strictly conserved
for the integrable Hamiltonian $H$, $[j_\text{E}, H] = 0$, the
staggered Zeeman term $H_\text{B}$ can be identified as the only
origin of scattering. This scattering can be treated perturbatively
according to Refs.\ \onlinecite{jung2006, jung2007, steinigeweg2010,
steinigeweg2011-2} if the strength of the magnetic field $B$ is a
sufficiently small parameter. In the time domain, we can formulate
such a perturbation theory in terms of the integro-differential
equation
\begin{equation}
\dot{C}_\text{E}(t) = - \int \limits_0^t \! \text{d}t' \, K(t-t') \,
C_\text{E}(t) \, , \label{master}
\end{equation}
where $K(t)$ is the memory kernel. To lowest order of $B$, $B^2$,
this memory kernel reads in the high-temperature limit $\beta \to 0$
\cite{steinigeweg2010, steinigeweg2011-2}
\begin{equation}
K(t) = \frac{\text{Tr} \{ \imath [J_\text{E}, H_B]_\text{I}(t) \,
\imath [J_\text{E}, H_B] \} }{\text{Tr} \{ J_\text{E}^2 \}} \propto
\frac{(B \, \Delta)^2}{1 + 2 \Delta^2} \, ,
\end{equation}
where the subscript $\text{I}$ of the first commutator indicates the
interaction picture w.r.t.\ $H$. Despite the integrability of $H$,
exactly calculating the time dependence of $K(t)$ is very difficult
and requires, e.g., the exact diagonalization of a finite system
\cite{jung2006, jung2007}. For our purposes, however, it is
sufficient to use the well-known Markov approximation $K(t) = K \,
\delta(t)$. This approximation is reasonable in the limit of small
magnetic fields $B \to 0$, where relaxation is arbitrarily slow. In
this way, we obtain from Eq.\ (\ref{master}) the exponential
relaxation
\begin{equation}
\frac{C_\text{E}(t)}{C_\text{E}(0)} = e^{-K \, t} \, , \quad
C_\text{E}(0) = \frac{1 + 2 \Delta^2}{32}
\end{equation}
corresponding to a Lorentzian line shape in frequency space. Thus,
using the expression for the dc conductivity $\kappa_\text{E}$ in
Eq.\ (\ref{TQ}) with $t_3 \to \infty$, we find the scaling
\begin{equation}
\kappa_\text{E}' = \frac{\kappa_\text{E}}{1 + 2 \Delta^2} \propto
\frac{1 + 2 \Delta^2}{(B \, \Delta)^2} \, .
\end{equation}
The quantity $\kappa_\text{E}'$ introduced does not include a
trivial scaling due to the sum rule $C_\text{E}(0)$ and is the
prediction of the perturbation theory as such. For small values of
$\Delta$, we eventually find the scaling $\kappa_\text{E} =
\kappa_\text{E}' \propto 1/(B \, \Delta)^2$ and, for large values of
$\Delta$, we find $\kappa_\text{E}'$ to be independent of $\Delta$.
In this case, $\kappa_\text{E} \propto (\Delta/B)^2$.

\begin{figure}[t]
\includegraphics[width=0.8\columnwidth]{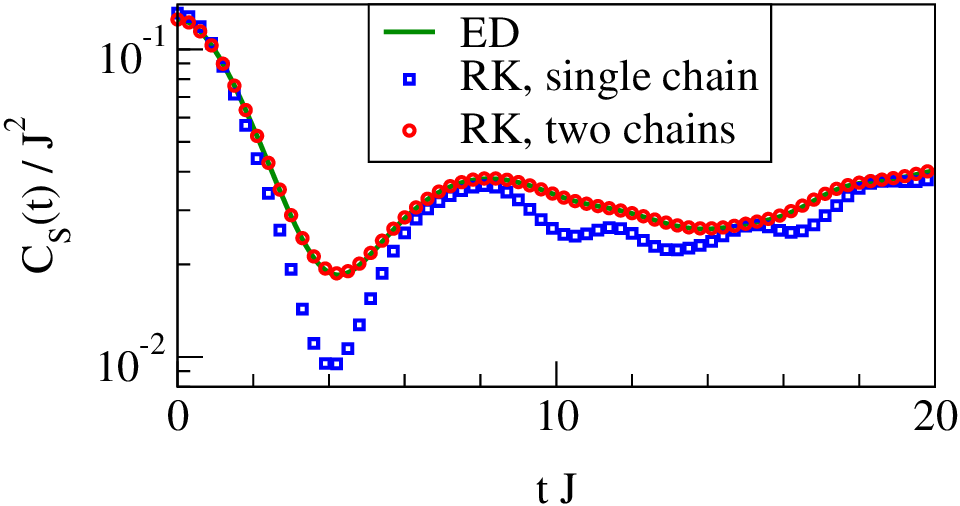}
\caption{(color online) Spin-current autocorrelation function
$C_\text{S}(t)$ in the integrable model at anisotropy $\Delta = 1.0$
and high temperatures $\beta \to 0$, numerically obtained for a
small system size $L=10$ using different methods: (i) exact
diagonalization, the approximation in Eq.\ (\ref{approximation2})
for a (ii) single chain and (iii) two identical, uncoupled chains.}
\label{Fig14}
\end{figure}

\section{Fourier Transform at High Temperatures}
\label{Fourier}

In Fig.\ \ref{Fig13} we show that the Fourier transform of our
numerical results for the time-dependent energy-current
autocorrelation function $C_\text{E}(t)$ at high temperatures, i.e.\
$\beta \to 0$, yields a Lorentzian line shape of
$C_\text{E}(\omega)$ in frequency space. This line shape is another
convincing indicator for the validity of conventional perturbation
theory.

\section{Typicality in Small Systems} \label{small}

Throughout this paper we have provided a comparison with
exact-diagonalization data to prove that typicality already holds in
finite systems of intermediate size, i.e., $L=14-18$. It is clear
that typicality has to break down for small systems. For the models
studied in this paper, we observe this breakdown for sizes below $L
\sim 10$, i.e., for effective dimensions below $d_\text{eff} \sim
1000$. In Fig.\ \ref{Fig14} we show one representative example.

However, a simple idea also allows us to use typicality for small
$L$: Consider two identical, uncoupled chains of length $L$ with the
Hamiltonian
\begin{equation}
H' = H \otimes 1 + 1 \otimes H
\end{equation}
and the current $j' = j \otimes 1 + 1 \otimes j$, respectively. In
this way, we do not change the exact current dynamics but increase
the dimension of the Hilbert space by a factor of $2^L$. For the
enlarged space, we can expect that typicality holds again. Figure
\ref{Fig14} verifies this expectation.

We emphasize that the preceding is equivalent to {\it averaging}.
The random state $|\psi\rangle$ from the $2^L \cdot 2^L$ dimensional
Hilbert space has a Schmidt decomposition
\begin{equation}
|\psi\rangle = \sum_{i=1}^{2^L} a_i \, | u_i \rangle \otimes | v_i
\rangle \, ,
\end{equation}
where $a_i$ are random coefficients and $|u_i\rangle$, $|v_i\rangle$
are random orthonormal basis sets for each of the $2^L$ dimensional
subspaces. Therefore, using a single random state in the doubled
system is equivalent to a randomly weighted average over $2 \cdot
2^L$ autocorrelations functions, each calculated using one of $2
\cdot 2^L$ randomly chosen states. If this is more efficient than
full diagonalization will depend on details.

We note that one can analogously consider more than two chains and
therefore use typicality for any smaller $L$.

\newpage


\begin{thebibliography}{99}

\bibitem{gemmer2003} J. Gemmer and G. Mahler,
Eur. Phys. J. B {\bf 31}, 249 (2003).

\bibitem{goldstein2006} S. Goldstein {\it et al.},
Phys. Rev. Lett. {\bf 96}, 050403 (2006).

\bibitem{popescu2006} S. Popescu, A. J. Short, and A. Winter,
Nature Phys. {\bf 2}, 754 (2006).

\bibitem{reimann2007} P. Reimann,
Phys. Rev. Lett. {\bf 99}, 160404 (2007).

\bibitem{white2009} S. R. White,
Phys. Rev. Lett. {\bf 102}, 190601 (2009).

\bibitem{sugiura2012}
S. Sugiura and A. Shimizu,
Phys. Rev. Lett. {\bf 108}, 240401 (2012).

\bibitem{bartsch2009} C. Bartsch and J. Gemmer,
Phys. Rev. Lett. {\bf 102}, 110403 (2009); EPL {\bf 96}, 60008 (2011).

\bibitem{elsayed2013} T. A. Elsayed and B. V. Fine,
Phys. Rev. Lett. {\bf 110}, 070404 (2013).

\bibitem{steinigeweg2014-1} R. Steinigeweg {\it et al.},
Phys. Rev. Lett. {\bf 112}, 130403 (2014).

\bibitem{steinigeweg2014-2} R. Steinigeweg, J. Gemmer, and W. Brenig,
Phys. Rev. Lett. {\bf 112}, 120601 (2014).

\bibitem{hams2000} A. Hams and H. De Raedt,
Phys. Rev. E {\bf 62}, 4365 (2000).

\bibitem{deutsch1991} J. M. Deutsch,
Phys. Rev. A {\bf 43}, 2046 (1991).

\bibitem{srednicki1994} M. Srednicki,
Phys. Rev. E {\bf 50}, 888 (1994).

\bibitem{rigol2008} M. Rigol, V. Dunjko, and M. Olshanii,
Nature {\bf 452}, 854 (2008).

\bibitem{cazalilla2010}
M. A. Cazalilla and M. Rigol, New J. Phys. {\bf 12}, 055006 (2010);
and references therein.

\bibitem{appelbaum2007}
I. Appelbaum, B. Huang, and D. J. Monsma,
Nature {\bf 447}, 295 (2007).

\bibitem{tombros2007}
N. Tombros {\it et al.},
Nature {\bf 448}, 571 (2007).

\bibitem{stern2008}
N. P. Stern {\it et al.},
Nature Phys. {\bf 4}, 843 (2008).

\bibitem{kuemmeth2008}
F. Kuemmeth {\it et al.},
Nature {\bf 452}, 448 (2008).

\bibitem{sologubenko2000} A. V. Sologubenko {\it et al.},
Phys. Rev. Lett. {\bf 84}, 2714 (2000).

\bibitem{hess2001} C. Hess {\it et al.},
Phys. Rev. B {\bf 64}, 184305 (2001).

\bibitem{hlubek2010} N. Hlubek {\it et al.},
Phys. Rev. B {\bf 81}, 20405R (2010).

\bibitem{thurber2001} K. R. Thurber {\it et al.},
Phys. Rev. Lett {\bf 87}, 247202 (2001).

\bibitem{kuehne2010} H. K\"{u}hne {\it et al.},
Phys. Rev. B {\bf 80}, 045110 (2009).

\bibitem{maeter2012} H. Maeter {\it et al.},
J. Phys.: Condens. Matter {\bf 25}, 365601 (2013).

\bibitem{xiao2014} F. Xiao {\it et al.},
arXiv:1406.3202 (2014).

\bibitem{zotos1997} X. Zotos, F. Naef, and Prelov\v{s}ek,
Phys. Rev. B {\bf 55}, 11029 (1997).

\bibitem{kluemper2002} A. Kl\"{umper} and K. Sakai,
J. Phys. A {\bf 35}, 2173 (2002).

\bibitem{shastry1990} B. S. Shastry and B. Sutherland,
Phys. Rev. Lett. {\bf 65}, 243 (1990).

\bibitem{zotos1999} X. Zotos,
Phys. Rev. Lett. {\bf 82}, 1764 (1999).

\bibitem{benz2005} J. Benz {\it et al.},
J. Phys. Soc. Jpn. {\bf 74}, 181 (2005).

\bibitem{prosen2011} T. Prosen,
Phys. Rev. Lett. {\bf 106}, 217206 (2011).

\bibitem{prosen2013} T. Prosen and E. Ilievski,
Phys. Rev. Lett. {\bf 111}, 057203 (2013).

\bibitem{narozhny1998} B. N. Narozhny, A. J. Millis, and N. Andrei,
Phys. Rev. B {\bf 58}, 2921R (1998).

\bibitem{heidrichmeisner2003} F. Heidrich-Meisner {\it et al.},
Phys. Rev. B {\bf 68}, 134436 (2003).

\bibitem{heidrichmeisner2007} F. Heidrich-Meisner, A. Honecker, and W. Brenig,
Eur. Phys. J. Special Topics {\bf 151}, 135 (2007).

\bibitem{herbrych2011} J. Herbrych, P. Prelov\v{s}ek, and X. Zotos,
Phys. Rev. B {\bf 84}, 155125 (2011).

\bibitem{steinigeweg2013} R. Steinigeweg, J. Herbrych, and P. Prelov\v{s}ek,
Phys. Rev. E {\bf 87}, 012118 (2013).

\bibitem{fujimoto2003} S. Fujimoto and N. Kawakami,
Phys. Rev. Lett. {\bf 90}, 197202 (2003).

\bibitem{johnston2000} D. C. Johnston {\it et al.},
Phys. Rev. B {\bf 61}, 9558 (2000).

\bibitem{trotzky2007} S. Trotzky {\it et al.},
Science {\bf 319}, 295 (2007).

\bibitem{gambardella2006} P. Gambardella,
Nature Mat. {\bf 5}, 431 (2006).

\bibitem{kruczenski2004} M. Kruczenski,
Phys. Rev. Lett. {\bf 93}, 161602 (2004).

\bibitem{kim1996} Y. B. Kim,
Phys. Rev. B {\bf 53}, 16420 (1996).

\bibitem{fabricius1998}
K. Fabricius and B. M. McCoy, Phys. Rev. B {\bf 57}, 8340 (1998).

\bibitem{steinigeweg2009} R. Steinigeweg and J. Gemmer,
Phys. Rev. B {\bf 80}, 184402 (2009).

\bibitem{steinigeweg2011-1} R. Steinigeweg and W. Brenig,
Phys. Rev. Lett. {\bf 107}, 250602 (2011).

\bibitem{prelovsek2013} A recent review is given in: P. Prelov\v{s}ek and J. Bon\v{c}a,
{\it Ground State and Finite Temperature Lanczos Methods} in
{\it Strongly Correlated Systems},
Solid-State Sciences {\bf 176} (Springer, Berlin, 2013).

\bibitem{mierzejewski2010} M. Mierzejewski and P. Prelov\v{s}ek,
Phys. Rev. Lett. {\bf 105}, 186405 (2010).

\bibitem{steinigeweg2012-1} R. Steinigeweg {\it et al.},
Phys. Rev. B {\bf 85}, 214409 (2012).

\bibitem{alvarez2002} J. V. Alvarez and C. Gros,
Phys. Rev. Lett. {\bf 88}, 077203 (2002).

\bibitem{grossjohann2010} S. Grossjohann and W. Brenig,
Phys. Rev. B {\bf 81}, 012404 (2010).

\bibitem{langer2009} S. Langer {\it et al.},
Phys. Rev. B {\bf 79}, 214409 (2009).

\bibitem{jesenko2011} S. Jesenko and M. \v{Z}nidari\v{c},
Phys. Rev. B {\bf 84}, 174438 (2011).

\bibitem{karrasch2012} C. Karrasch, J. H. Bardarson, and J. E. Moore,
Phys. Rev. Lett. {\bf 108}, 227206 (2012).

\bibitem{karrasch2013-1} C. Karrasch {\it et al.},
Phys. Rev. B {\bf 87}, 245128 (2013).

\bibitem{karrasch2014-1} C. Karrasch, J. E. Moore, and F. Heidrich-Meisner,
Phys. Rev. B {\bf 89}, 075139 (2014).

\bibitem{karrasch2014-2} C. Karrasch, D. M. Kennes, J. E. Moore,
Phys. Rev. B {\bf 90}, 155104 (2014).

\bibitem{huang2013} Y. Huang, C. Karrasch, and J. E. Moore,
Phys. Rev. B {\bf 88}, 115126 (2013).

\bibitem{karrasch2013-2} C. Karrasch, R. Ilan, and J. E. Moore,
Phys. Rev. B {\bf 88}, 195129 (2013).

\bibitem{prosen2009} T. Prosen and M. \v{Z}nidari\v{c},
J. Stat. Mech.: Theory Exp. {\bf 2009}, P02035.

\bibitem{znidaric2011} M. \v{Z}nidari\v{c},
Phys. Rev. Lett. {\bf 106}, 220601 (2011).

\bibitem{alcantarabonfim1992} O. F. de Alcantara Bonfim and G. Reiter,
Phys. Rev. Lett. {\bf 69}, 367 (1992); Phys. Rev. Lett. {\bf 70},
249 (1993).

\bibitem{gerling1993} R. W. Gerling and H. Leschke,
Phys. Rev. Lett. {\bf 70}, 248 (1993).

\bibitem{steinigeweg2012-2} R. Steinigeweg,
EPL {\bf 97}, 67001 (2012).

\bibitem{sirker2009} J. Sirker, R. G. Pereira, and I. Affleck,
Phys. Rev. Lett. {\bf 103}, 216602 (2009); Phys. Rev. B {\bf 83},
035115 (2011).

\bibitem{steinigeweg2010} R. Steinigeweg and R. Schnalle,
Phys. Rev. E {\bf 82}, 040103R (2010).

\bibitem{steinigeweg2011-2} R. Steinigeweg,
Phys. Rev. E {\bf 84}, 011136 (2011).

\bibitem{jung2006} P. Jung, R. W. Helmes, and A. Rosch,
Phys. Rev. Lett. {\bf 96}, 067202 (2006).

\bibitem{jung2007} P. Jung and A. Rosch,
Phys. Rev. B {\bf 76}, 245108 (2007).

\bibitem{deraedt2007} K. De Raedt {\it et al.},
Comp. Phys. Comm. {\bf 176}, 121 (2007).

\bibitem{jin2010} F. Jin {\it et al.},
J. Phys. Soc. Jpn. {\bf 79}, 124005 (2010).

\bibitem{kolezhuk2004} A. Kolezhuk and H. Mikeska,
Lect. Not. Phys. {\bf 645}, 1 (2004).

\bibitem{kubo1991} R. Kubo, M. Toda, and N. Hashitsume,
{\it Statistical Physics II: Nonequilibrium Statistical Mechanics}
(Springer, Berlin, 1991).

\bibitem{carmelo2014} J. M. P. Carmelo, T. Prosen, and D. K. Campbell,
preprint, arXiv:1407.0732 (2014).

\bibitem{evenodd} For $\Delta=0$, the XXZ Hamiltonian describes a
model of free spin-less fermions, $H = J \sum_{i=1}^L \cos(2 \pi
i/L) \, n_i$, where the single-particle dispersion consists of only
$L$ points at positions $2 \pi i/L$. Thus, for finite $L$, the
spectrum of any correlation function in the vicinity of $\Delta=0$
is necessarily sparse and, except for trivial cases, has to feature
strong finite-size effects, e.g., even-odd effects.

\end{thebibliography}
\end{document}